\RequirePackage{fix-cm}
\documentclass[smallcondensed]{svjour3}       %
\smartqed  %
\usepackage[authoryear]{natbib}
\usepackage{graphicx}
\usepackage{hyperref}
\usepackage{xcolor}
\usepackage{multirow}
\usepackage[most]{tcolorbox}

\definecolor{darkgreen}{rgb}{0.0, 0.5, 0.0}

\usepackage{graphicx}
\usepackage{booktabs}
\usepackage[most]{tcolorbox}
\usepackage{listings} 
\usepackage{graphicx} %
\usepackage{xcolor}
\usepackage[inkscapelatex=false]{svg}
\usepackage{svg}
\usepackage{xurl}
\svgpath{{../figures/}}
\usepackage{enumitem}
\usepackage{acronym}

\usepackage{parskip} %
\usepackage{hyperref} %

\definecolor{boxgray}{rgb}{1, 1, 1}
\definecolor{darkermutedcoral}{RGB}{233, 116, 81}
\definecolor{vividsoftblue}{RGB}{70, 130, 180}
\definecolor{softteal}{RGB}{72, 164, 156}
\definecolor{black}{RGB}{0, 0, 0}

\lstdefinestyle{mystyle}{
    backgroundcolor=\color{white},          %
    commentstyle=\color{darkermutedcoral},  %
    keywordstyle=\color{vividsoftblue},     %
    numberstyle=\tiny\color{black},         %
    stringstyle=\color{softteal},           %
    basicstyle=\ttfamily\tiny\color{black}, %
    breakatwhitespace=false,         
    breaklines=true,                 
    captionpos=b,                    
    keepspaces=true,                 
    numbers=left,                    
    numbersep=5pt,                  
    showspaces=false,                
    showstringspaces=false,
    showtabs=false,                  
    tabsize=2
}

\newcommand{\rqone}{\textit{What are the characteristics of the output formats across LLMs and prompts?}}
\newcommand{\rqtwo}{\textit{To what extent can the output format of LLMs be controlled using prompt engineering and lightweight post-processing?}}
\newcommand{\rqthree}{\textit{What's the impact of output control on the reported performance of LLMs in terms of CA?}}
\newcommand{\rqfour}{\textit{Does inconsistent output format introduce bias in text-oriented metric-based evaluation?}}

\newcommand{\rqfive}{\textit{Does output format bias exist in the evaluation of closed LLMs?}}

\acrodef{llm}[LLM]{Large Language Model}
\acrodefplural{llm}[LLMs]{Large Language Models}
\acrodef{pl}[PL]{programming language}
\acrodefplural{pl}[PLs]{programming languages}
\acrodef{ca}[CA]{Computational Accuracy}
\acrodef{csr}[CSR]{Code Extraction Success Rate}
\acrodef{msr}[MSR]{Match Success Rate}
\acrodef{cr}[CR]{Compilation Rate}
\acrodef{ase}[ASE]{automated software engineering}

\begin{document}

\title{Output Format Biases in the Evaluation of Large Language Models for Code Translation}

\titlerunning{Output Format Biases in the Evaluation of LLMs for Code Translation}

\author{Marcos Macedo       \and
        Yuan Tian   \and
        Filipe R. Cogo \and
        Bram Adams
}

\institute{Marcos Macedo, Yuan Tian, Bram Adams\at
              School of Computing, Queen's University, ON, Canada \\
             \email{marcos.macedo@queensu.ca, y.tian@queensu.ca, bram.adams@queensu.ca}
           \and
           Filipe R. Cogo \at
               Centre for Software Excellence, Huawei Canada \\
              \email{filipe.roseiro.cogo1@huawei.com}
}

\date{Received: date / Accepted: date}

\maketitle

\begin{abstract}
Code translation between \acp{pl} is a long-existing and critical task in software engineering, facilitating the modernization of legacy systems, ensuring cross-platform compatibility, and enhancing software performance. With the recent advances in \acp{llm} and their applications to code translation, there is an increasing need for comprehensive evaluation of these models. Most existing studies instruct \acp{llm} to perform code translation and evaluate their performance by either running the generated outputs through test suites or comparing them to reference outputs (ground truth). These outputs, however, may contain not only executable source code but also additional non-code elements, such as natural language explanations or formatting tokens. We refer to the way source code and non-code elements are combined as \textit{output format}. It is crucial to understand and address variations in output format, as non-code elements can interfere with evaluation metrics, resulting in biased (inaccurate or unfair) assessments of model performance and comparisons. We refer to this bias as \textit{output format bias}. To investigate the presence of output format biases, we first conduct an empirical analysis of the outputs from eleven instruct-tuned open-source LLMs, applied to 3,820 translation pairs across five languages: C, C++, Go, Java, and Python. The results show that between 26.4\% and 73.7\% of outputs produced by our evaluated \acp{llm} necessitate post-processing (to result in source code for evaluation). To mitigate output format bias, we propose a strategic combination of prompt engineering and regular expressions that effectively extracts source code from mixed-format outputs, enabling the eleven open-source models to achieve an average Code Extraction Success Rate (CSR) of 92.73\%. Our empirical study confirms that output format bias affects widely used execution-based metrics, i.e., Computational Accuracy (CA), and text-based metrics, i.e., BLEU, CodeBLEU and CrystalBLEU. Additionally, we test five closed \acp{llm} and observe that they also generate varying distributions of output formats, which could contribute to output format biases. Our results highlight the need to mitigate the output format bias to enable reliable evaluations in \acp{llm} code translation.

\end{abstract}

\maketitle

\section{Introduction} \label{sec:intro}
Automated code translation, i.e., translating source code between different programming languages (\acp{pl}), promises to save software development teams substantial time and effort while minimizing the risks associated with manual translation errors and inconsistencies. As more companies seek to migrate their existing software systems from outdated PLs to more contemporary ones or adapt their software for cloud-based environments where specific PLs are better suited, the value of automated code translation becomes increasingly apparent~\citep{weisz_perfection_2021}.

Traditionally, code translation has been implemented using transpilers, such as C2Rust\footnote{\url{https://github.com/immunant/c2rust}} and cxgo\footnote{\url{https://github.com/gotranspile/cxgo}}, which rely on hard-coded rules and program analysis to convert code between languages. These tools are costly to develop and limited to specific language pairs. Furthermore, they often produce translations that do not align with the idiomatic expressions of the target language~\citep{szafraniec_code_2023}. Recently, pre-trained models have gained prominence in code translation~\citep{roziere2020unsupervised,lachaux2021dobf,roziere2021leveraging,szafraniec_code_2023}. These models are pre-trained on large-scale source code datasets and occasionally fine-tuned on code translation pairs. \acp{llm} offer new opportunities in code translation by enabling prompt engineering, an alternative to traditional model parameter adjustments, which leverages \emph{prompts} to guide \acp{llm} in addressing diverse tasks. The introduction of \acp{llm} and prompt engineering have significantly advanced \ac{ase} tasks, elevating their performance to new records~\citep{fan2023large}.

Interestingly, despite the advancements of \ac{llm}-based approaches in different \ac{ase} tasks, a recent study by Pan et al.~\citeyear{pan2024lost} shows that \acp{llm} are yet to prove their reliability to automate code translation, with the rate of translations that can completely pass the accompanying unit test(s), i.e., \ac{ca}~\citep{roziere2020unsupervised}, ranging from 2.1\% to 47.3\% on average. Pan et al. manually categorized the bugs introduced by \acp{llm} in code translation and found that most (77.8\%) unsuccessful translations result in compilation errors. In this paper, we postulate that the practical evaluation of \acp{llm} for code translation revolves around the models' capability to generate correctly translated code. Nonetheless, generated translations deemed as invalid during \ac{llm} evaluation (e.g., due to compilation errors) often stem from the problem of \emph{inconsistent output format}, not necessarily from the \acp{llm}' limitations for code translation. Inconsistent output format refers to a problem where the translated source code is presented in the generated output in various formats and could be interspersed with natural language (non-source-code) (ref. Fig.~\ref{fig:output-formats}). The \textbf{output format bias} occurs when variations in how the output is structured (output format), rather than the accuracy of the translation, affect the perceived quality of the translation result. Therefore, it is important to make researchers and practitioners aware of the impact of inconsistent output format on the evaluations of \acp{llm} for code translation.  By distinguishing between errors caused by output format bias and those stemming from the quality of the translated code, we can more reliably assess the ability of \acp{llm} to translate source code between different \acp{pl}.

The output format bias can influence the calculation of many important benchmark metrics for evaluating code translation such as:

\begin{figure}[t]
\centering
    \begin{tcbitemize}[raster columns=2,raster equal height, raster width=.95\linewidth, colback=white!5!white,colframe=white!45!black]
    \centering
    \tcbitem[squeezed title*={(a) Direct Output}, fonttitle=\small]
        \begin{lstlisting}[language=Python]
def main():
    a, b = input("Enter two numbers: ").split()
    a, b = int(a), int(b)
    if a*b%
        print("Even")
    else:
        print("Odd")

### Explanation:

The Python code is similar to the Go code, with a few differences in syntax.
        \end{lstlisting}
    \tcbitem[squeezed title*={(b) Wrapped Code}, fonttitle=\small]
        \begin{lstlisting}[language=C]
```c
#include <stdio.h>
#include <string.h>

int main() {
    char input[100];
    char doubleInput[200];
    ...
    } else {
        printf("Yes\n");
    }
    return 0;
}
```

This C code does ...
        \end{lstlisting}
    \tcbitem[squeezed title*={(c) Unbalanced Back-ticks}, fonttitle=\small]
        \begin{lstlisting}[language=Python]
#!/usr/bin/env python
N, D = map(int, input().split())

ans = N // (D2+1)
rem = N %

if rem != 0:
    ans += 1

print(ans)
```
        \end{lstlisting}
    \tcbitem[squeezed title*={(d) Re-Wrapped Code}, fonttitle=\small]
        \begin{lstlisting}[language=C++]
```
Here is the C++ equivalent [...] code:

```cpp
#include <iostream>
using namespace std;

int main() {
    int A, B, K;
    cin >> A >> B >> K;
    ...
    return 0;
}
```

This C++ code performs ...
        \end{lstlisting}
    \end{tcbitemize}
\caption{Examples of the observed output formats in our study. \textbf{(a)} Python code in Direct Output with Additional text (the explanation). \textbf{(b)} C code wrapped with three back-ticks followed by the language extension, with Additional text. \textbf{(c)} Python code that has three back-ticks at the end but no matching opening back-ticks. \textbf{(d)} This output format occurs in RQ2 as the model does not generate code after the back-ticks and instead generates a completely new code block. Some code examples are shortened for brevity.}
\label{fig:output-formats}
\end{figure}

\begin{itemize}[leftmargin=*]
    \item \textbf{Text-oriented metrics}: Inconsistent output formats lead to wrong estimates of text-oriented evaluation metrics, as generated text other than translated code, such as code explanations, is accounted for in their calculations. These include BLEU~\citep{papineni_bleu_2001}, CodeBLEU~\citep{ren_codebleu_2020} and CrystalBLEU~\citep{eghbali2022crystalbleu}. These metrics measure either the overlap of tokens or the semantic similarity between a translation and its reference. 
    \item \textbf{Execution-based metrics}: Inconsistent output formats lead to wrong estimates when calculating execution-based metrics such as Computational Accuracy (CA) \cite{roziere2021leveraging}. While the code excerpt in the output can be a potentially valid translation of the reference source code, the whole output fails compilation. This leads to the conclusion that the model could not generate the correct translated code when, in fact, the problem stems from the generated natural language text, which does not comply with the source code syntax.
\end{itemize}

To validate our hypothesis on the prevalence of output format bias and explore whether it can be mitigated through cost-effective techniques, such as prompt engineering and lightweight post-processing, we conduct a case study analyzing the outputs of 16 popular \acp{llm} (ref. Table~\ref{table:models}) for code translation across multiple programming languages. We first focus on eleven open-source models, which are more likely to deviate from instructions and generate additional text, and then expand the analysis to five closed models. Our investigation spans 3,820 code translation tasks, with source code samples drawn from the well-known CodeNet benchmark~\citep{puri_codenet_2021}, and target code generated by the \acp{llm}. The dataset includes 20 translation pairs across five programming languages (C, C++, Go, Java, and Python). Specifically, we address the following five research questions (RQs) in this case study:

\vspace{1mm} \noindent \textbf{RQ1 \rqone} We analyze the output formats generated by eleven open-source \acp{llm} and find that these models generate outputs in three distinct source code formats, some of which include natural language (additional text) and some that do not. Of these six possible combinations (output format with or without additional text), only one is directly parsable without further processing.

\vspace{1mm} \noindent \textbf{RQ2 \rqtwo} We verify that combining prompt instructions to control the output format with a regular expression for code extraction can increase the proportion of source code extracted from the inference output by up to 41.8\%. 

\vspace{1mm} \noindent \textbf{RQ3 \rqthree} We use different combinations of prompt and extraction methods and study their impact on the most widely adopted execution-based evaluation metric, i.e., the Computational Accuracy (CA) metric. The proposed lightweight approach presents an average \ac{ca} up to 6 times higher than a direct compilation.

\vspace{1mm} \noindent \textbf{RQ4 \rqfour} We analyze the impact of the output format on text-oriented metrics, specifically BLEU \citep{papineni_bleu_2001}, CodeBLEU \citep{ren_codebleu_2020} and CrystalBLEU \citep{eghbali2022crystalbleu}. Our results show that applying our proposed lightweight approach (Regex) to extract source code leads to improvements in BLEU scores in up to 72.08\% of cases and in CrystalBLEU scores in up to 50.56\% of cases, when additional text is present in the output. Furthermore, we observe that the presence of additional text artificially inflates the Data Flow component of CodeBLEU, leading to counterintuitive outcomes, such as a decrease in CodeBLEU scores when additional text is present alongside the source code. %

\vspace{1mm} \noindent \textbf{RQ5 \rqfive} We investigate whether the problem of multiple output formats also affects closed-source models by manually examining the inference outputs of five closed \acp{llm}: Gemini 1.5 Flash, GPT-4o, GPT-4o mini, GPT-4, and GPT-3.5 Turbo. Our analysis reveals that while some models have a consistent output format, some of them present a distribution, similar to the open-source ones. For instance, Gemini 1.5 Flash, GPT-4o, and GPT-4o mini consistently produce outputs exclusively in the Wrapped-Code format, allowing regular expressions to achieve a 100\% \ac{csr}. In contrast, GPT-3.5 Turbo and GPT-4 primarily generate outputs in the Direct Output format, with only a small fraction (1.67\%) in the Wrapped-Code format. %

This work extends our prior conference paper~\citep{macedo2024exploring}, which initiated the investigation into the impact of output format on \ac{llm}-based code translation evaluation. The original paper focused on output format bias in open-source LLMs and examined it using a single execution-based metric (i.e., CA). In this extended study, we address two additional research questions: how output format inconsistency affects the calculation of two text-oriented metrics (RQ4) and whether closed \ac{llm} also experiences inconsistent output formats (RQ5). Additionally, we conduct an ablation study on AlphaTrans \citep{ibrahimzada_alphatrans_2025}, a recent repository-level code translation approach that incorporates our proposed output format mitigation method, to better understand the impact of this mitigation strategy, as discussed in Section~\ref{sec:mitigation-in-practice}. %

In summary, the main contributions of this article are as follows:
\begin{itemize}[leftmargin=*]
    \item We show how often open-source and closed \acp{llm} consistently output the expected code format for benchmark evaluation in code translation.
    \item We demonstrate a cost-effective output control method to increase the chances of reliably retrieving the source code from the generation output models and calculate the method's effectiveness for code translation.
    \item We assess the impact of output control on the reported performance of \acp{llm} in code translation in terms of compilation rate and \ac{ca}.
    \item We constructed a new dataset with human-written ground truth translations by augmenting the Pan et al. dataset with accepted solutions from the CodeNet repository, enabling the use of text-based evaluation metrics such as BLEU, CodeBLEU and CrystalBLEU.
    \item We show the influence of natural language in evaluating code translation using text-oriented metrics. 
    \item We demonstrate the significant practical impact of addressing output format bias through an ablation study on a state-of-the-art, repository-level code translation system.
    \item To foster future research in the area, we have made our replication package publicly available on GitHub\footnote{\url{https://github.com/RISElabQueens/forge24-code-translation}}.
\end{itemize}

This paper is organized as follows. Section~\ref{sec:background} provides background about literature work and code translation evaluation. Section~\ref{sec:method} presents our study's design and implementation details. In Sections \ref{sec:rq1}--\ref{sec:rq5}, we present the motivation, approach, and results for each of the five raised research questions. We discuss the implications and suggestions from our study in Section~\ref{sec:disc} and threats to validity in Section~\ref{sec:threats}. Finally, we conclude in Section~\ref{sec:conclusion}.

\section{Background and Related Work}\label{sec:background}
This section introduces background and related work on automated code translation (Section~\ref{sec:related_code_translation}), benchmark evaluation (Section~\ref{sec:related_benchmarks}), and prompt engineering (Section~\ref{sec:related_prompt}).

\subsection{Automated Code Translation}\label{sec:related_code_translation}

Automated code translation is a long-existing task in software engineering (SE). Existing approaches can be categorized into four classes, rule-based (e.g., C2Rust\footnote{\url{https://github.com/immunant/c2rust}}, statistical learning-based~\citep{nguyen2013lexical,nguyen2014migrating,karaivanov2014phrase}, neural network-based~\citep{chen2018tree}, and pre-trained model-based~\citep{lachaux2021dobf,roziere2021leveraging,szafraniec_code_2023, jiao2023evaluation, pan2024lost, yang2024exploring}. The first category encompasses tools that utilize program analysis techniques and handcrafted rules to translate code from one \ac{pl} to another. For instance, C2Rust and cxgo\footnote{\url{https://github.com/gotranspile/cxgo}} are designed to convert C programs into Rust and Go, respectively. Sharpen\footnote{\url{https://github.com/mono/sharpen}} and Java2CSharp\footnote{\url{https://github.com/bdqnghi/j2cstranslator}} are tools for transforming Java code into C\#. While these tools are employed in the industry, their development can be costly. Additionally, they often result in translations that may not align with the idiomatic usage of the target language, leading to challenges in readability for programmers~\citep{szafraniec_code_2023}. As representatives for the second category, Nguyen et al.~\citeyear{nguyen2013lexical, nguyen2014migrating},  Karaivanov et al.~\citeyear{karaivanov2014phrase} and Aggarwal et al.~\citeyear{aggarwal2015using} investigated how well statistical machine translation models for natural languages could apply to code translation. These methods overlook the grammatical structures inherent in \acp{pl}. To mitigate this issue, Chen et al.~\citeyear{chen2018tree} proposed a tree-to-tree neural network that translates a source tree into a target tree by using an attention mechanism to guide the expansion of the decoder. Their approach outperforms other neural translation models designed for translating human languages.

Pre-trained model-based approaches leverage self-supervised learning to train a model from large-scale source code for code translation. Roziere et al.~\citeyear{roziere2020unsupervised} introduced an unsupervised code translation model that only requires source code in multiple \acp{pl} without parallel code translation pairs, which are often hard to collect. They developed TransCoder, a model pre-trained on GitHub's open-source projects leveraging three tasks, i.e., masked language modeling, recovering corrected code, and back-translation. Building on this, they later introduced DOBF~\citep{lachaux2021dobf} and TransCoder-ST~\citep{roziere2021leveraging}. DOBF is a model pre-trained to reverse code obfuscation by employing a sequence-to-sequence learning model. TransCoder-ST, on the other hand, utilizes automatic test generation techniques to automatically select and fine-tune high-quality translation pairs for the pre-trained model. Szafraniec et al. proposed TransCoder-IR~\citep{szafraniec_code_2023}, enhancing TransCoder by incorporating low-level compiler intermediate representations. In addition to models specifically pre-trained for code translation, general-purpose pre-trained models for source code, such as CodeT5~\citep{wang2021codet5} and CodeBERT~\citep{feng2020codebert}, have also shown promise in code translation tasks. These models can be effectively fine-tuned on parallel code translation pairs, as demonstrated in the work of Jiao et al.~\citeyear{jiao2023evaluation}.

Recently, \acp{llm} have gained prominence as highly effective tools for code intelligence tasks, including code translation~\citep{fan2023large,yang2024exploring}. Different from previous pre-trained model-based approaches, \acp{llm} are designed to understand and follow human instructions, enabling their straightforward application in various downstream tasks through prompt-based interactions rather than expensive task-specific fine-tuning or pre-training. Pan et al.~\citeyear{pan2024lost} investigate the effectiveness of \acp{llm} in code translation. The authors collected executable code samples from various datasets and projects and performed translations using popular \acp{llm} including GPT-4~\citep{achiam2023gpt}, Llama 2~\citeyear{LLama-2}, Wizard Vicuna 13B~\citeyear{Wizard-Vicuna-13B}, Airoboros 13B~\citeyear{airoboros-13B}, StarCoder~\citep{li_starcoder_2023}, CodeGeeX~\citep{zheng2023codegeex}, and CodeGen~\citep{nijkamp2022codegen}. They reported that \acp{llm} were largely ineffective in translating real-world projects, highlighting the need for improvement in code translation techniques. A more recent study by Yang et al.~\citeyear{yang2024exploring} reported similar findings on the errors generated by LLM-based code translation approaches. They propose UniTrans, a method that generates test cases for the target program and guides LLMs to iteratively repair incorrectly translated code based on the test case execution results.

\subsection{Code Translation Benchmarks and Evaluation Metrics} \label{sec:related_benchmarks} 
Lu et al.~\citeyear{lu2021codexglue} unveiled CodeXGLUE, a benchmark dataset designed to assess machine learning models across ten different program understanding and generation tasks. This dataset, compiled from a variety of open-source projects, includes a specialized code-to-code translation dataset featuring 11,800 Java to C\# translation pairs. In a similar vein, Puri et al.~\citeyear{puri_codenet_2021} introduced CodeNet, a diverse dataset that covers 55 programming languages. CodeNet primarily focuses on code translation tasks, with its data sourced from two prominent online judge platforms: AIZU\footnote{\url{https://onlinejudge.u-aizu.ac.jp/}} and AtCoder\footnote{\url{https://atcoder.jp/}}.

Zhu et al.~\citeyear{zhu2022multilingual} contributed to this growing field with CoST, a parallel code corpus that encompasses seven languages and facilitates snippet-level alignment through code comment matching. CoST is versatile, supporting a range of tasks, including code translation, summarization, and synthesis. Building upon CoST, Zhu et al. further introduced XLCoST~\citep{zhu2022xlcost}, a dataset offering alignment at both snippet and program levels. Both CoST and XLCoST were curated from GeeksForGeeks\footnote{\url{https://www.geeksforgeeks.org/}}, a resource-rich website featuring a plethora of data structures and algorithm problems, along with solutions in up to seven popular programming languages.

More recently, Jiao et al.~\citeyear{jiao2023evaluation} developed G-TransEval, a benchmark that integrates various existing benchmarks into a unique benchmark by categorizing code translation pairs into four types: syntax level, library level, semantic level, and algorithm level translations. 

When utilizing the benchmarks mentioned above, the evaluation of code translation approaches typically relies on two types of metrics: text-oriented and execution-based. Text-oriented metrics, such as BLEU~\citep{papineni_bleu_2001}, CodeBLEU~\citep{ren_codebleu_2020}, and CrystalBLEU~\citep{eghbali2022crystalbleu}, measure token overlap or semantic similarity between the generated code and the reference. However, these metrics can be skewed by inconsistent output formats that include non-code elements like explanations. Execution-based metrics, such as Computational Accuracy (CA)~\citep{roziere2021leveraging}, assess whether the generated code compiles and runs correctly. Inconsistent formats, where non-code elements cause compilation failures, can lead to inaccurate conclusions about model performance, even when the code itself is valid. Note that text-oriented metrics require a parallel dataset (i.e., paired translation tasks with human-written reference translations), which is not always provided in the benchmarks mentioned above and may require additional steps to prepare.

\subsection{Prompt Engineering in SE} \label{sec:related_prompt}

The remarkable ability of \acp{llm} for zero-shot and few-shot prompting has gained growing interest in exploring how in-context learning and prompt engineering can improve \ac{llm}-based applications within SE. Researchers have leveraged \acp{llm} and prompt engineering for various \ac{ase} tasks~\citep{feng2023prompting,pan2024lost,gao2023constructing,li2023nuances,geng2024large}. 

Gao et al.~\citeyear{gao2023constructing} conducted a study to understand the impact of the choice, order, and number of demonstration examples on the effectiveness of in-context learning in code intelligence tasks. Pan et al.~\citeyear{pan2024lost} introduced an iterative prompting method that integrates additional contextual information to improve \ac{llm}-based code translation. This method enriches the input to language models with comprehensive context, including the original code, prior prompts, erroneous translations, detailed error information, translation instructions, and expected outcomes. Yang et al.~\citeyear{yang2024exploring} follows a similar iterative approach with additional test cases generated. Li et al.~\citeyear{li2023nuances} explored ChatGPT's capability in identifying test cases that reveal bugs in source code. Initially, ChatGPT showed limited success, but with careful prompting, its performance improved significantly. Feng et al.~\citeyear{feng2023prompting} utilized prompts that exploit few-shot learning and chain-of-thought reasoning for \ac{llm}-based bug reproduction. Geng et al.~\citeyear{geng2024large} delved into the efficacy of \acp{llm} in generating code comments with various intents, focusing on their attributes. They employed the in-context learning approach and designed new strategies for selecting and re-ranking examples to optimize performance.

\section{Study Setup} \label{sec:method}
This section discusses the selection of \acp{llm} for our study (Section~\ref{sec:method:subsec:llm}), the data collection procedure (Section~\ref{sec:method:subsec:data}), and the utilized prompt templates (Section~\ref{sec:method:subsec:prompt}).

\subsection{Selected Large Language Models}\label{sec:method:subsec:llm}

A diverse range of models exists in the rapidly evolving field of \acp{llm} for code generation, offering a variety of capabilities and performance. In this case study, we select 16 targets \acp{llm}, of which 11 are open-sourced, while the remaining five are closed. For all models, we select instruct-tuned \acp{llm} to the detriment of their associated base models, as instruct-tuned models are fine-tuned to follow prompted instructions more effectively. Table~\ref{table:models} summarizes the basic information of selected \acp{llm}.

\begin{table}[!ht]
\centering
\caption{Summary of \acp{llm} used in the current study. }
\resizebox{\linewidth}{!}{%
\begin{tabular}{llrrl}
\toprule
\textbf{Source Type} & \textbf{Model Name} & \textbf{Size (Billions)} & \textbf{Context Length (Tokens)} & \textbf{Release Date} \\
\midrule
\multirow{11}{*}{Open-source} 
& CodeLlama Instruct & 7 & 16,384 & Aug, 2023 \\
& CodeLlama Instruct & 13 & 16,384 & Aug, 2023 \\
& CodeLlama Instruct & 34 & 16,384 & Aug, 2023 \\
& Magicoder-S-CL & 7 & 16,384 & Dec, 2023 \\
& Magicoder-CL & 7 & 16,384 & Dec, 2023 \\
& WizardCoder & 1 & 8,192 & Aug, 2023 \\
& WizardCoder & 3 & 8,192 & Aug, 2023 \\
& WizardCoder Python & 7 & 8,192 & Aug, 2023 \\
& WizardCoder Python & 13 & 8,192 & Aug, 2023 \\
& WizardCoder Python & 34 & 8,192 & Aug, 2023 \\
& Mixtral 8x7B Instruct v0.1 & 46.7 & 8,192 & Dec, 2023 \\
\midrule
\multirow{5}{*}{Closed} 
& GPT-3.5 Turbo & Unspecified & 16,384 & Nov, 2023 \\
& GPT-4 & Unspecified & 8,192 & Mar, 2023 \\
& GPT-4o & Unspecified & 128,000 & May, 2024 \\
& GPT-4o mini & Unspecified & 128,000 & Jul, 2024 \\
& Gemini 1.5 Flash & Unspecified & 1,000,000 & Feb, 2024 \\
\bottomrule
\end{tabular}
}
\label{table:models}
\end{table}

These models can be grouped across six distinct model families: 
\begin{itemize}[leftmargin=*]
    \item (Open-source) Magicoder~\citep{wei_magicoder_2023}: This collection of \acp{llm}, trained on 75K synthetic instruction data using the OSS-Instruct framework, includes two variants: Magicoder-CL and Magicoder-S-CL, optimized for code generation and understanding.
    
    \item (Open-source) WizardCoder~\citep{luo_wizardcoder_2023}: Built on the StarCoder framework~\citep{li_starcoder_2023}, WizardCoder specializes in generating complex code instruction data. It comes in various sizes, ranging from 1 billion to 34 billion parameters, and is known for its strong performance on intricate coding tasks.
    
    \item (Open-source) CodeLlama~\citep{roziere_code_2023}: A family of models derived from Llama 2, designed specifically for code generation tasks. CodeLlama models are available in three sizes: 7B, 13B, and 34B parameters, providing flexibility for different levels of computational resources and code complexity.
    
    \item (Open-source) Mixtral~\citeyear{mistralai2023mixtral}: A Mixture-of-Experts model developed by Mistral AI, known for its robust performance in code generation. \textit{Note: While Mixtral has both open and closed variants, we specifically use the open-source version in this study.}
    
    \item (Closed) OpenAI's GPT\footnote{\url{https://platform.openai.com/docs/models}}: A family of models developed by OpenAI, specifically GPT-3.5 Turbo, GPT-4, GPT-4o, and GPT-4o Mini. These models differ in their size, context lengths, and use cases. GPT-3.5 Turbo is a lightweight model offering fast responses with moderate capabilities, while GPT-4 and GPT-4o provide superior reasoning, handling more complex tasks. GPT-4o Mini, a cost-efficient version of GPT-4o, is designed for large-scale and time-sensitive applications.
    
    \item (Closed) Gemini\footnote{\url{https://ai.google.dev/gemini-api/docs/models/gemini}}: Specifically, Gemini 1.5 Flash. It is the most recent LLM released by Google with a massive 1 million token context window. 
\end{itemize}

When selecting open-source \acp{llm}, we consider two key factors, i.e., they demonstrated strong performance in code generation tasks, and they align well with efficient deployment frameworks, such as vLLM~\citep{kwon_efficient_2023}, without precluding their compatibility with other platforms like the HuggingFace Text Generation Interface~\citep{wolf_huggingfaces_2020}. This consideration ensures that our chosen models are not only high-performing but also practically feasible for widespread use in research and industry applications.

For closed \acp{llm}, their selection is motivated by their commercial relevance and state-of-the-art performance. Models like GPT-4 and Gemini 1.5 Flash are leading commercial solutions that demonstrate exceptional capabilities in handling complex code generation tasks at scale. Their widespread use in industry highlights their robustness, while their closed nature allows them to integrate advanced proprietary optimizations that open models may not yet achieve. Studying these models enables us to benchmark against the very best available technologies and understand their advantages in practical applications. We also include older models, such as GPT-3.5 Turbo, for comparison purposes, as their preferences in output format may differ from those of more recent models, potentially providing valuable insights into the evolution of format generation across different versions.

\subsection{Dataset and Preprocessing}\label{sec:method:subsec:data}

\noindent \textbf{Preparing Data for RQ1, RQ3, and RQ5:} We utilize programs from the dataset provided by Pan et al.~\citeyear{pan2024lost} that are derived from the CodeNet dataset. 
This dataset contains 1,000 programs in five \acp{pl} (C, C++, Go, Java, and Python), with 200 programs and test cases for each programming language.

There are 20 possible translation combinations of source-target program language pairs considering the five \acp{pl} found in the dataset. 
Therefore, we create 4,000 translation samples that consist of programs in the source programming language and the 4 desired output languages (e.g., one program in C can be translated to C++, Go, Java, and Python). %

\begin{figure}[ht]
\centering
\includegraphics[width=0.8\textwidth]{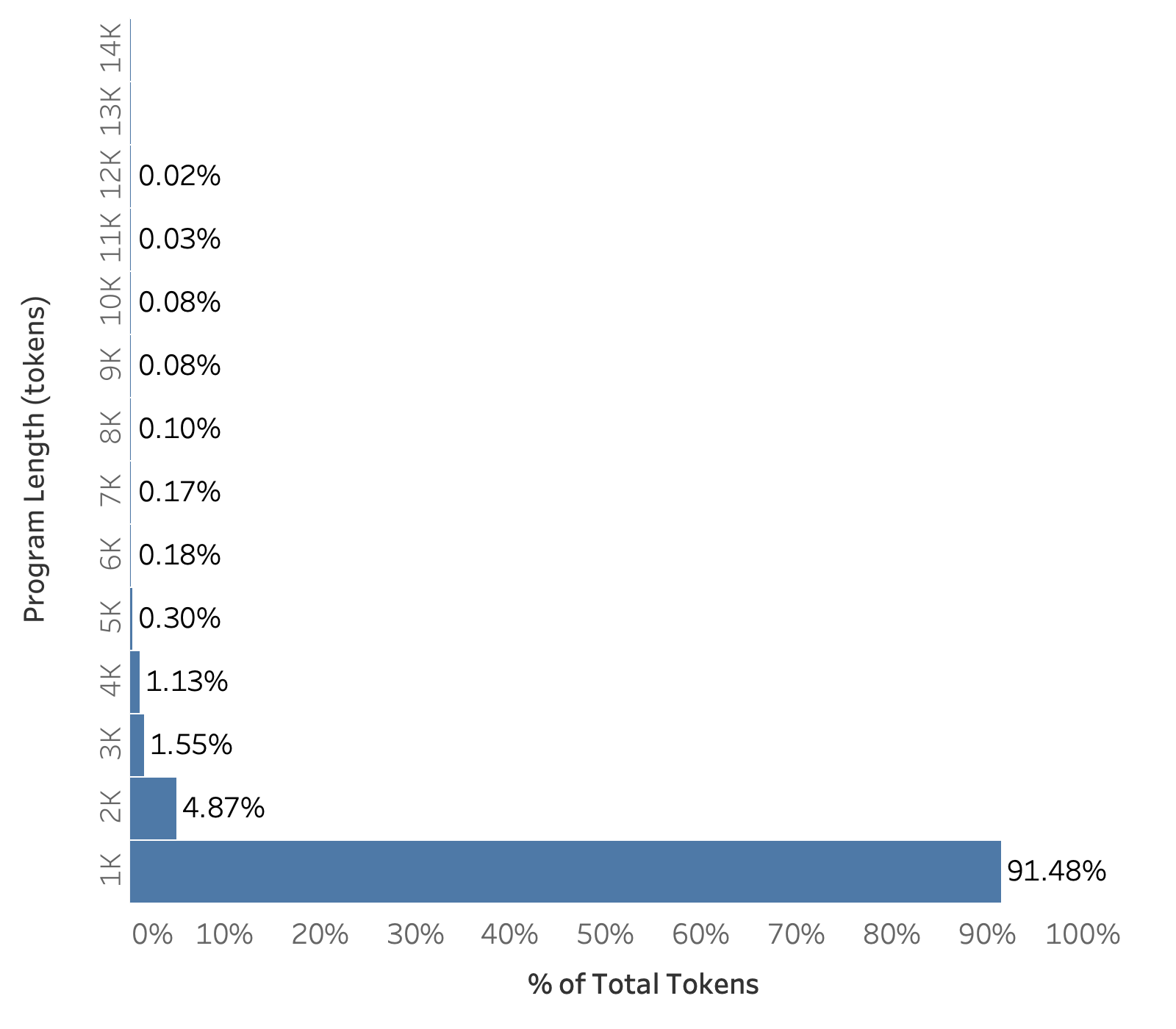}
\caption{Distribution of program length in Pan et al.'s dataset before the cut-off. }

\label{fig:code-distribution-before}
\end{figure}

We use the tokenizer of each of the eleven models to tokenize the input code (program) from each of the code pairs.
We can observe in Fig.~\ref{fig:code-distribution-before} that 97.9\% of the programs have a source code token length of up to 3,072 tokens.
Therefore, we utilize this number as the cut-off for code pairs, filtering out from our dataset code pairs that have an input program token length greater than 3,072 tokens. In addition, we allow models to generate up to 2,048 new tokens during inference. Doing so ensures that we can fit the input code, prompt, and output into the context window of all models.

\begin{figure}[ht]
\centering
\includegraphics[width=0.85\textwidth]{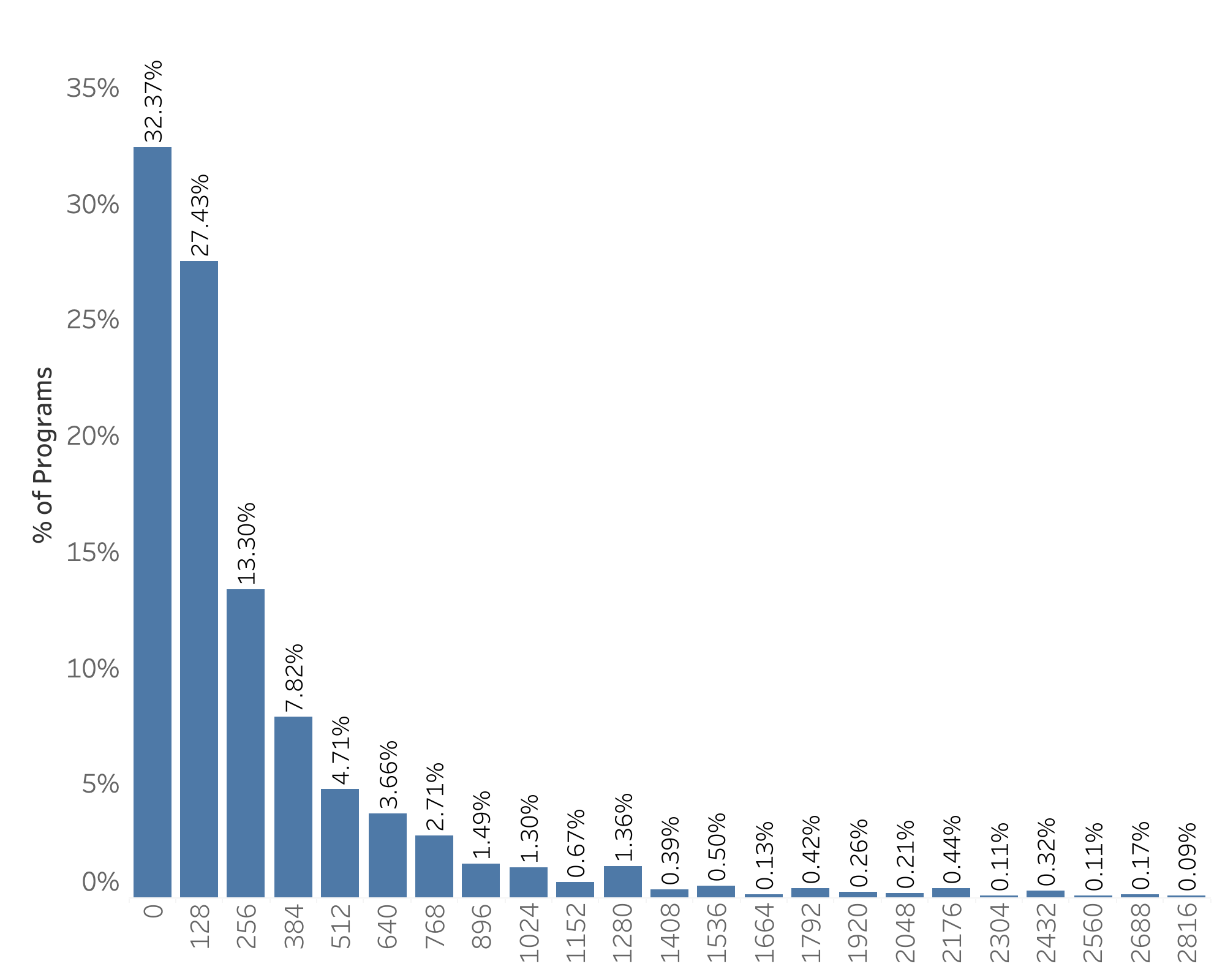}
\caption{Distribution of the token lengths of programs in our dataset.}
\label{fig:token-distribution-after}
\end{figure}

After excluding input code exceeding 3,072 tokens in length, our dataset was reduced to 3,912 translation samples.
However, the filtering process resulted in an imbalanced dataset due to varying numbers of code pairs for each source-target language combination. 
As a result, the number of code translation samples across some PL pairs is greater than that of other PL pairs. We then balance the number of code pairs in each source-target language combination by down-sampling those with a higher frequency.
In the end, we have a dataset of 3,820 code pairs, with 191 code pairs and test cases for each of the 20 source-target language combinations.
Fig.~\ref{fig:token-distribution-after} showcases the distribution of the final dataset.

\vspace{0.1cm}
\noindent \textbf{Preparing Data for RQ4:} The dataset from Pan et al. lacks human-written ground truth translations for each sample across \acp{pl} (i.e., parallel data), which is essential for computing text-oriented metrics like CodeBLEU and BLEU required to address RQ4. This absence poses a challenge in evaluating translation quality. To overcome this, we searched the CodeNet repository for corresponding source programs in our filtered dataset, consisting of 3,820 translation pairs. Once we identified the problem solved by the source program in CodeNet, we selected an accepted solution in the target language to serve as the ground truth for comparison.

Since not all problems have accepted solutions available in every target language, we excluded these cases from the dataset to maintain the consistency and relevance of our analysis. After this refinement, our final dataset retained 93.17\% of the original filtered samples, resulting in 3,657 translation pairs for answering RQ4.

\subsection{Prompt Templates}\label{sec:method:subsec:prompt}

\begin{figure*}[h]
\centering
\includegraphics[width=\textwidth]{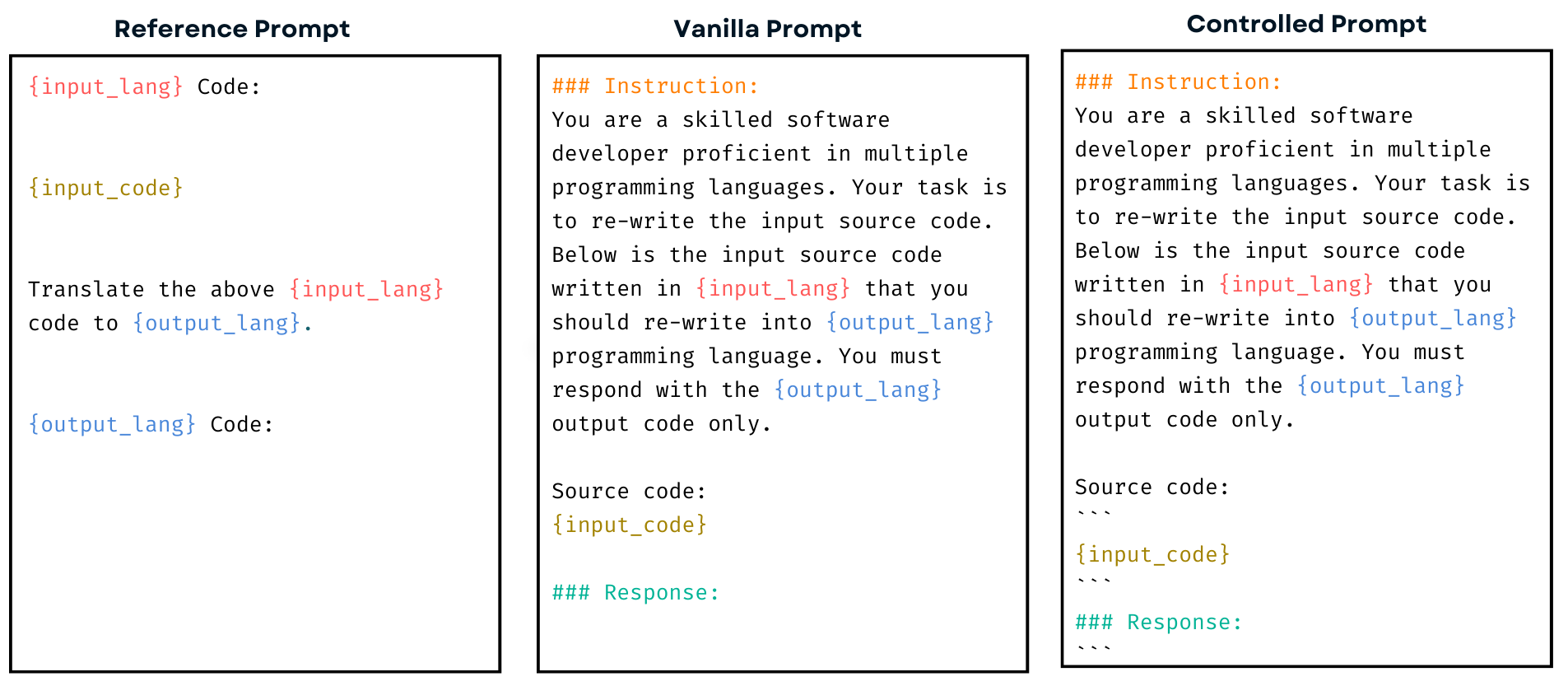}
\caption{Examples of the Prompt Templates used in WizardCoder.
The same instruction is used for all the \acp{llm}, following the recommended prompt template provided by the authors of each model, as outlined in their respective model cards. The exception is Reference Prompt, which is used as-is.}
\label{fig:prompts-used}
\end{figure*}

We use three prompt templates in our case study, as seen in Fig.~\ref{fig:prompts-used}.
The first template is proposed by Pan et al.~\citeyear{pan2024lost} for evaluating open-source \acp{llm}. 
We refer to it as ``Reference Prompt''. 
The ``Vanilla Prompt'' prompt template, created by us, is tailored for each model based on the model's recommended template from its respective paper or HuggingFace model card (e.g., WizardCoder \footnote{\url{https://huggingface.co/WizardLMTeam/WizardCoder-15B-V1.0}}).%

Additionally, for the instruction in our prompt template, we follow OpenAI's suggested prompting strategy, i.e., ``Ask the model to adopt a persona''~\footnote{\url{https://platform.openai.com/docs/guides/prompt-engineering/strategy-give-models-time-to-think}}. This strategy is also adopted in recent instruct-tuned code \acp{llm}, such as Magicoder~\citep{wei_magicoder_2023}.

Lastly, \textit{Controlled Prompt} is a variation of \textit{Vanilla Prompt} designed to control the output format of the models. 
The detailed design concerns for this prompt can be found in Section~\ref{sec:rq2}.

\subsection{Evaluation Metrics}\label{sec:method:subsec:metric} %

In this study, we consider both widely used execution-based evaluation metrics and text-oriented metrics. Specifically, we considered the following four evaluation-based metrics when comparing selected \acp{llm}.

\begin{itemize}
    \item \textbf{Computational Accuracy (CA):} Rozière et al.~\citeyear{roziere2020unsupervised} introduced the \ac{ca} metric, which evaluates whether a transformed target function generates the same outputs as the source function when given the same inputs. The target function is correct if it gives the same output as the source function for every tested input value. In our case, the programs in the dataset are functions that receive input from \texttt{stdin} and produce an output.
    \item \textbf{Compilation Rate (CR)}: The number of programs that successfully compiled over the total number of programs in the datasets.
    \item \textbf{Match Success Rate (MSR)}: The number of generated outputs where the regular expression designed to extract source code successfully finds a match in the output, divided by the total number of generated outputs.
    \item \textbf{Code Extraction Success Rate (CSR)}: The number of generated outputs for which we are able to extract source code, divided by the total number of generated outputs. 
\end{itemize}
MSR and CSR are new evaluation metrics that we developed based on our proposed method for controlling the output format, which are explained in Section~\ref{sec:rq2}. Note that, ``match'' is a pre-condition to ``extract'' the source code from the output.

For text-oriented metrics, we consider BLEU and CodeBLEU: 
\begin{itemize}
\item \textbf{BLEU}: The BLEU metric, originally introduced by Papineni et al.~(\citeyear{papineni_bleu_2001}), measures the overlap between the n-grams of the machine-generated translation and the reference (human-written) translation. It is computed as a geometric mean of n-gram precision, combined with a brevity penalty to penalize overly short translations. In the context of code translation, BLEU evaluates how closely the generated code matches the reference code at a token level, where tokens are typically keywords, operators, or identifiers in the code.
\item \textbf{CodeBLEU}: Ren et al.~(\citeyear{ren_codebleu_2020}) proposed CodeBLEU as an enhancement of BLEU specifically designed for code generation tasks. In addition to n-gram precision, CodeBLEU incorporates structural matching (comparing code abstract syntax trees), data flow, and keyword matching to better account for the unique features of programming languages. It combines the traditional BLEU score with these code-specific properties, yielding a more accurate reflection of code quality and functional correctness.
\item \textbf{CrystalBLEU}: Proposed by Eghbali et al.~(\citeyear{eghbali2022crystalbleu}), this metric builds on the strengths of BLEU while addressing its susceptibility to noise from trivially shared n-grams. These are common n-grams that frequently appear across code in the same language but do not indicate meaningful similarity. By filtering out these trivial overlaps, CrystalBLEU enhances \textit{distinguishability}, a property defined by the authors to quantify how much more similar semantically equivalent code examples are compared to non-equivalent ones.

\end{itemize}

\subsection{Implementation Details}\label{sec:method:subsec:implementation}

We begin by downloading the official checkpoints for all evaluated models from Hugging Face. %
Subsequently, the vLLM framework is employed to conduct inference using Greedy Decoding on four Nvidia RTX 6000 GPUs. Using Greedy Decoding guarantees the reproducibility of the inference outputs across multiple runs. For models ranging from 1 to 7B in size, a single GPU is utilized. Models with 13B employ Tensor Parallelism across two GPUs, while those exceeding 13B leverage the same technique across all four GPUs. In the case of closed source models, we use their inference API with greedy decoding. Regarding the execution and compilation of programs, our environment is equipped with the following versions of compilers and run-times: OpenJDK 11, Python 3.11, g++ 12, gcc 12, and Go 1.19.

\section{RQ1: \rqone} \label{sec:rq1}

Examining the output formats generated by the bench-marked models is a crucial aspect of validating our hypothesis. 
Additionally, this evaluation sheds light on the prevalent output formats across various models and inspires the development of approaches to control the output format.

\subsection{Approach}
\label{sec:rq1:subsec:approach}

Our initial step involved creating a representative sample of our dataset to examine the distribution and variety of output formats produced by these models. This was achieved through stratified sampling
across source-target \ac{pl} combinations within the dataset. As a result, we acquired a subset of 360 code translation pairs, spanning 20 source-target language pairs, which accurately represent our dataset.

For each of the 11 open-source models, we utilized both the Reference and Vanilla Prompt templates (as illustrated in Fig.~\ref{fig:prompts-used}) to generate inference results. With each model producing 360 inference outcomes, the total number of inferences across all models amounted to 3,960 for each prompt. 

To identify and summarize patterns within these outputs, we conducted open-coding as guided by the method described by Stol et al.~\citep{stol_grounded_2016}. We further extracted a representative subset of these output formats for each prompt to ensure a comprehensive analysis. This involved stratified sampling across the language pairs and \acp{llm}, leading to a subset of 440 entries that reflects the broader set of 3,960 results.

Finally, we estimated the proportions of these identified patterns within the inference results generated from our dataset. This process not only helps in understanding the behavior of different \acp{llm} in response to varying prompts, but also contributes to the broader understanding of \ac{llm} output formats in the context of code translations.

\subsection{Results}
\label{sec:rq1:subsec:results}

We analyze the output formats generated by models in response to two different prompt templates, i.e., the Reference and Vanilla Prompt templates. The models produce outputs in three primary formats in terms of the source code:

\begin{enumerate}[leftmargin=*]
    \item \textbf{Direct Output}: This format consists solely of source code without any special formatting.
    \item \textbf{Wrapped Code}: Here, the source code is enclosed within a code block, delineated by triple back-ticks.
    \item \textbf{Unbalanced Back-ticks}: In this format, the source code is followed by a closing triple back-tick, but lacks an opening one.
\end{enumerate}

For each of these output formats, we further categorize the results based on the presence or absence of natural language elements, such as notes, comments, or explanations, accompanying the source code:

\begin{itemize}[leftmargin=*]
    \item \textbf{No additional text}: The inference results contain only source code, with no additional natural language content.
    \item \textbf{Additional text}: Natural language is present in the inference results. This may appear before, after, or interspersed within the code.
\end{itemize}

Fig.~\ref{fig:output-formats} illustrates three sample outputs from our dataset, along with their corresponding sub-categories.

\begin{figure}[ht]
\centering
\includegraphics[width=0.8\textwidth]{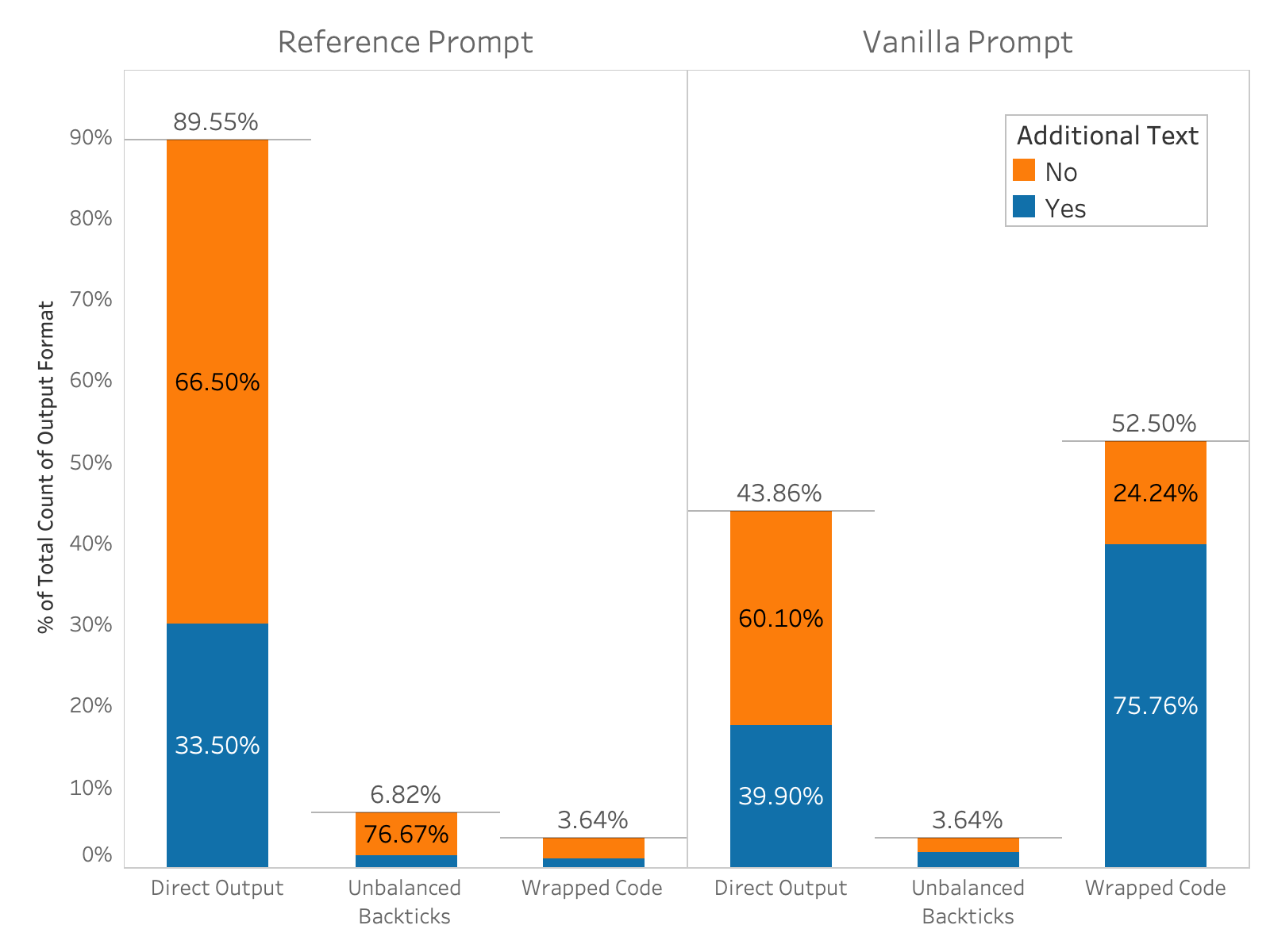}
\caption{Distribution of output formats observed for each prompt in RQ1. The height of the bar represents the observed proportion of output formats over the sampled generation outputs for the prompt.} 
\label{fig:plot-rq1}
\end{figure}

\noindent {\textbf{Finding 1. The models' output format is not consistent across our dataset, and post-processing is required to extract the source code in up to 73.64\% of the outputs.} Among the six combinations (three source code formats, with or without \textit{Additional text}), only when the models generate the \textit{Direct Output} format without \textit{Additional text}, the inference output can be directly employed for evaluation. This is because it solely consists of source code and thus necessitates no further post-processing. As shown in Fig.~\ref{fig:plot-rq1}, in the case of our \textit{Vanilla Prompt}, only 26.36\% (43.86\% $\times$ 60.10\%) of the generated outputs contain solely source code and are immediately usable (over the total 43.86\% that are Direct Output, we observe that 60.10\% have no additional text, as seen in Fig.~\ref{fig:plot-rq1}). The remaining 73.64\% (100\% - 26.36\%) of source code requires post-processing and can not be directly extracted due to added natural text such as explanations, comments, and notes, or being in a different output format than what we expect (Wrapped Code, Unbalanced Back-Ticks).

Conversely, for our Reference Prompt, we estimate that up to $59.5$\% (89.55\% $\times$ 66.50\%) of the outputs are suitable for direct evaluation without post-processing. This percentage represents the outputs in \textit{Direct Code Output} format that do not contain additional text. However, the remaining 40.5\% of the outputs will cause compilation (or interpretation) errors due to added comments or being in other output formats than the assumed one.

\noindent \textbf{Finding 2. Different prompts produce different output format distributions}
As shown in Fig.~\ref{fig:plot-rq1}, the proportions of the three categories of output formats vary based on the design of the prompt. In total, 89.55\% of the outputs by the Reference Prompt have a Direct Code format. On the other hand, the Vanilla Prompt outputs mostly \textit{Wrapped Code} format (in 52.50\% of the inferences). This result shows that not only do the models generate different amounts of \textit{Additional Text} alongside source code in the Direct Output category (as observed in Finding 1), but also the predominant category of output format varies across prompts. 

\noindent \textbf{Finding 3. During inference, models disregard the instruction to \textit{output code only} 59.3\% of the time.}
\textit{Vanilla Prompt} instruct the models to "Output code only" as seen in Fig. ~\ref{fig:prompts-used}. We find that in 59.32\% of the cases across all the output formats, the model added comments alongside the source code (as derived from Fig. ~\ref{fig:plot-rq1}). Surprisingly, this number is higher than the \textit{Reference Prompt} that does not instruct the model to output only code. In the latter, only 32.73\% of the generated output across all output formats have additional text alongside the source code.

\begin{tcolorbox}[enhanced,width=\textwidth,size=fbox,drop shadow southwest,sharp corners]

\textit{RQ1 Summary:} We observe that the models generate outputs in varying formats and distributions, depending on the prompt used. In extensive studies comparing different models or prompts, this variation in output formats presents a challenge in understanding the distribution of possible output formats and accurately capturing code. This variation significantly affects the reliability of the metrics derived from the models' generated output.
\end{tcolorbox}

\section{RQ2: \rqtwo} \label{sec:rq2}
In RQ1 (Section~\ref{sec:rq1}), we confirm that \acp{llm} can generate outputs with different formats. Thus, we aim to explore the extent to which the output format of \acp{llm} can be controlled (i.e., shifting the output distribution to one preferred type) so that we can extract code automatically in a unified way.

\subsection{Approach}
Our approach to control the models' output format uses a combination of prompt engineering and regular expression (regex) parsing.

Our prompt engineering method attempts to improve the consistency of the output format across all the models, such that the generated code in the output can be directly extracted for the highest number of programs possible without capturing comments that would interfere with the compilation or interpretation of the code. For prompt engineering, we modify the \textit{Vanilla Prompt} (Section~\ref{sec:method:subsec:prompt}) by adding a control statement that instructs the model to generate the code excerpt within triple back-ticks (Wrapped Code format). Fig.~\ref{fig:prompts-used} (c) shows the derived \textit{Controlled Prompt}.

We use the Regex in Fig.~\ref{fig:regex} to find and extract source code from the generated outputs. The Regex is designed to match the \textit{Wrapped Code} format (i.e., the text inside the first block delimited by three back-ticks on the output). The regex also ignores, if present, the first line in the output that often denotes the programming language the code is written into as it can be seen in Fig.~\ref{fig:output-formats}.

\begin{figure}[ht]
\centering
\begin{lstlisting}[basicstyle=\ttfamily\footnotesize, breaklines=true, numbers=none]
Response:?.*?```(?:(?:java|cpp|csharp|python|c|go|C\+\+|Java|Python|C#|C|Go))?(.+)```
\end{lstlisting}
\caption{Regex designed to match and extract source code from code blocks (Wrapped Code format) in the generated output text.}
\label{fig:regex}
\end{figure}

We obtain a subset of 440 samples through stratified sampling across the \ac{pl} pairs and \acp{llm}, following the same methodology described in RQ1 (Section~\ref{sec:rq1:subsec:approach}) and estimating the proportion of each of the output formats. We estimate our Regex's accuracy (\ac{msr}) on this subset of 440 samples. For the cases where there was a match, we analyze the percentage of matched source code.
We also manually analyze the cases with no match and explain the reasons for such mismatches.

\subsection{Results}
Fig.~\ref{fig:plot-rq2} shows the distribution of output format types between the \textit{Vanilla Prompt} and the \textit{Controlled Prompt}.

\begin{figure}[h]
\centering
\includegraphics[width=0.84\textwidth]{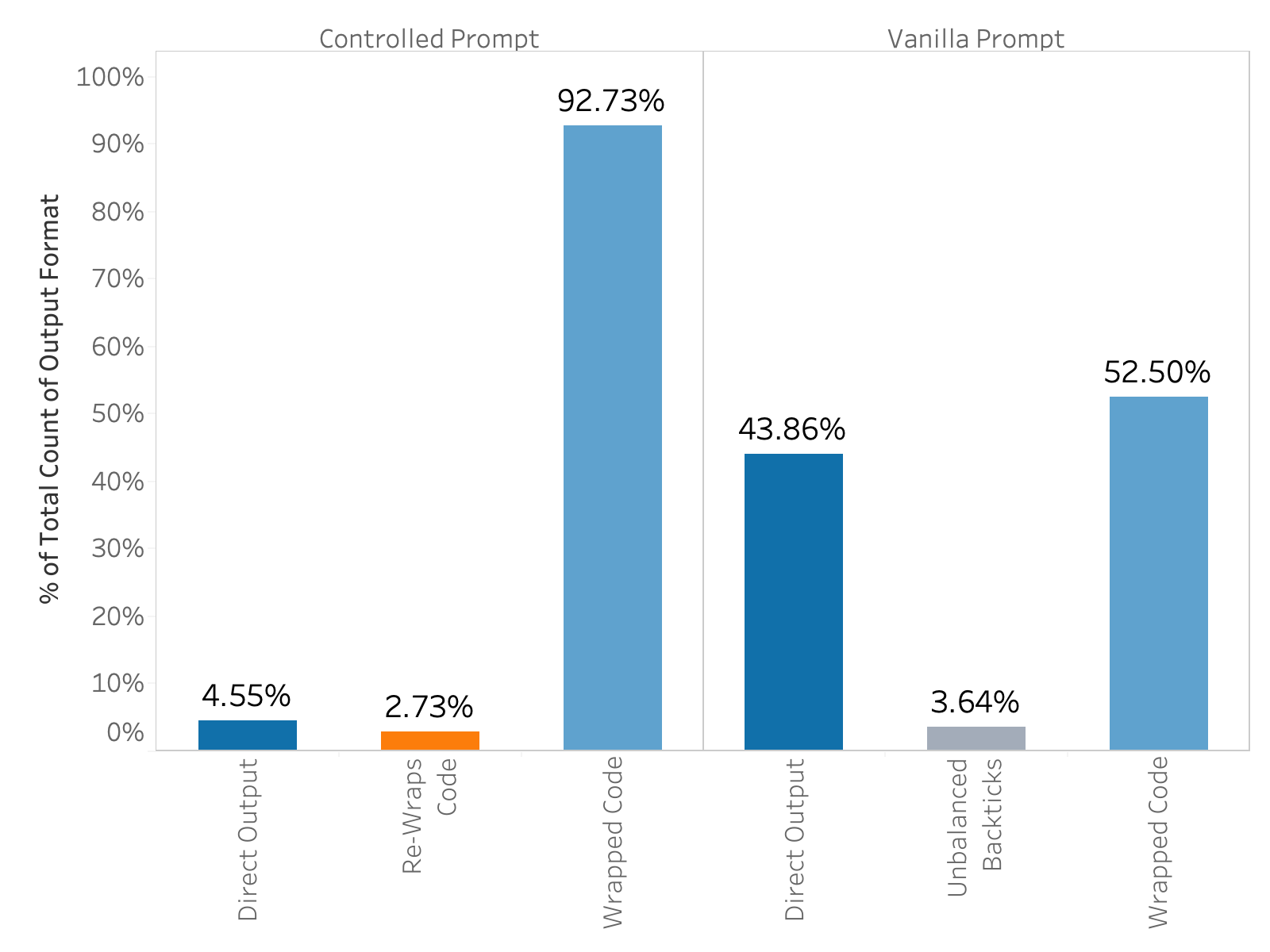}
\caption{Comparison of the distribution of output formats of the Controlled Prompt (left side) vs Vanilla Prompt (right side).} %
\label{fig:plot-rq2}
\end{figure}

\noindent \textbf{Finding 4. The output format can be controlled through a combination of prompt engineering and lightweight post-processing, achieving a \ac{msr} of 95.45\% and a \ac{csr} of 92.73\%.} Our proposed prompt is capable of steering the output across models, and our proposed Regex (Fig.~\ref{fig:regex}) matches 95.45\% of the generated outputs. As the matched content could be source code with added text, we manually examine the matches to confirm what proportion of the matches are source code only.
We observe that out of the 95.45\% there is 2.73\% of matches that are not source code. In other words, we achieve a \ac{csr} of 92.73\%. 

Next, we perform a detailed analysis on the 2.73\% of matches that do not contain source code. Through this analysis, we discovered that the content matched comprised natural language text rather than source code. More interestingly, 100\% of those cases are caused by the Mixtral 8x7B model. Specifically, we observe the emergence of a new output format we name \textbf{Re-Wraps Code} generated by the Mixtral 8x7B model, as seen in Fig.~\ref{fig:output-formats} (d). In the \textit{Re-Wraps Code} format, the model does not complete the code block opened using back-ticks in the \textit{Controlled Prompt}, ignoring our output control mechanism. Instead, the model opens a new set of back-ticks, adds source code and finishes by closing the back-tick group. Because our Regex is designed to capture \textit{Wrapped Code}, when it is applied to the Re-Wrapped Code format it matches from the beginning of the back-ticks we introduced in the \textit{Controlled Prompt} up to the first set of back-ticks generated by the model, not capturing the source code.

\begin{tcolorbox}[enhanced,width=\textwidth,size=fbox,drop shadow southwest,sharp corners]

\textit{RQ2 Summary:} The output format can be controlled to increase the percentage of output that can be matched by the Regex from 52.50\% in the Vanilla Prompt to 95.45\% in the Controlled Prompt. Controlling the output significantly helps capture more source code. \end{tcolorbox}

\vspace{0.1cm}

\section{RQ3: \rqthree} \label{sec:rq3}
In RQ2, we show the feasibility of controlling output formats of \acp{llm} for code translation via a combination of prompt engineering and lightweight post-processing regex. As the \ac{msr} changes, we hypothesize that commonly considered evaluation metrics for code translation (i.e., \ac{cr} and \ac{ca}) should also change. Thus, in this RQ we empirically quantify what is the impact of our output control on the reported performances of \acp{llm} on the whole 3,820 translation pairs.

\subsection{Approach}
To answer this question, we leverage the \textit{Vanilla Prompt} and \textit{Controlled Prompt} that are shown in Fig.~\ref{fig:prompts-used}. Using each of the prompts, we perform translation on all the 3,820 code pairs in our dataset. We employ various combinations of prompts and source code extraction methods to explore their effects on the results reported by three execution-based metrics. The evaluated combinations are:

\begin{enumerate}[leftmargin=*]
  \item \textbf{Vanilla + Direct Evaluation (VDE):} We utilize the Vanilla Prompt and directly evaluate the generated output. We do not perform any regex matching or source code extraction.
  \item \textbf{Vanilla + Regex (VRE):} We utilize the Vanilla Prompt and apply our regex to extract source code.
  \item \textbf{Controlled + Regex (CRE):} We utilize the Controlled Prompt and apply our regex to extract source code.
\end{enumerate}

\subsection{Results}
Table~\ref{tab:combined-models} shows the \ac{msr}, \ac{cr}, and \ac{ca} for the eleven considered models using three considered scenarios (VDE, VRE, and CRE) on the 3,820 code translation pairs.

\begin{table*}[ht]
\centering
\caption{Results for different combinations of prompt and extraction method across eleven LLMs. VDE refers to (Vanilla + Direct Evaluation), VRE refers to (Vanilla + Regex), and CRE refers to (Controlled + Regex) combinations.} 
\label{tab:combined-models}
\resizebox{\textwidth}{!}{%
\begin{tabular}{@{}lrrrrrrrrrrrrrr@{}}
\toprule
& & \multicolumn{3}{l}{\textbf{Match Success Rate (MSR)}} & \multicolumn{3}{l}{\textbf{Compilation Rate (CR)}} & \multicolumn{3}{l}{\textbf{Average CA}} \\
\cmidrule(lr){3-5} \cmidrule(lr){6-8} \cmidrule(lr){9-11}
\textbf{Model Name} & \multicolumn{1}{l}{\textbf{Model Size}} & \multicolumn{1}{l}{\textbf{CRE}} & \multicolumn{1}{l}{\textbf{VRE}} & \multicolumn{1}{l}{\textbf{VDE}} & \multicolumn{1}{l}{\textbf{CRE}} & \multicolumn{1}{l}{\textbf{VRE}} & \multicolumn{1}{l}{\textbf{VDE}} & \multicolumn{1}{l}{\textbf{CRE}} & \multicolumn{1}{l}{\textbf{VRE}} & \multicolumn{1}{l}{\textbf{VDE}} \\
\midrule
CodeLlama Instruct & 7   & 94.2\% & 6.60\%  & -- & 50.42\% & 2.77\%   & \textbf{30.03\%} & 33.27\% & 1.83\%   & 18.30\% \\
CodeLlama Instruct & 13  & 96.9\% & 0.68\%  & -- & 57.70\% & 0.05\%   & 27.20\% & 38.12\% & 0\%      & \textbf{18.51\%} \\
CodeLlama Instruct & 34  & 98.1\% & 40.16\% & -- & 49.29\% & 22.80\%  & 7.93\% & 34.06\% & 13.17\%  & 5.18\% \\
Magicoder-CL          & 7   & 88.5\% & 67.98\% & -- & 56.47\% & 47.20\%  & 0\% & 37.33\% & 32.02\%  & 0\% \\
Magicoder-S-CL        & 7   & 99.0\% & \textbf{99.35\%} & -- & \textbf{68.09\%} & \textbf{68.80\%}  & 0\% & 43.30\% & \textbf{43.12\%}  & 0\% \\
Mixtral 8x7B          & 46.7 & 60.4\% & 82.20\% & -- & 0\%     & 53.51\%  & 2.17\% & 0\%     & 35.60\%  & 1.28\% \\
WizardCoder           & 1   & 96.8\% & 72.91\% & -- & 47.39\% & 48.72\%  & 0.13\% & 18.61\% & 17.25\% & 0.03\% \\
WizardCoder           & 3   & 96.9\% & 68.09\% & -- & 59.45\% & 53.74\%  & 3.90\% & 32.28\% & 29.92\% & 0.55\% \\
WizardCoder Python    & 7   & 98.3\% & 40.65\% & -- & 55.29\% & 29.35\%  & 15.79\% & 34.11\% & 15.79\% & 10.13\% \\
WizardCoder Python    & 13  & 98.9\% & 71.28\% & -- & 62.93\% & 47.80\%  & 0.31\% & 36.20\% & 27.64\% & 0.1\% \\
WizardCoder Python    & 34  & \textbf{99.4\%} & 28.48\% & -- & 66.49\% & 18.38\%  & 0\% & \textbf{43.85\%} & 15.42\% & 0\% \\
\midrule
\textbf{Average} & \textbf{--} & \textbf{93.40\%} & \textbf{52.58\%} & \textbf{--} & \textbf{52.19\%} & \textbf{35.74\%} & \textbf{7.95\%} & \textbf{31.92\%} & \textbf{20.98\%} & \textbf{4.92\%} \\
\bottomrule
\end{tabular}
}
\end{table*}

\vspace{0.1cm}
\noindent \textbf{Finding 5: The integration of prompt engineering with a lightweight post-processing technique (CRE) demonstrated superior \ac{msr}, \ac{cr}, and \ac{ca} among the tested combinations.} The CRE approach achieved an impressive average \ac{msr} of 93.40\%, coupled with an average \ac{cr} of 52.19\% and an average \ac{ca} of 35.74\% over the models.

The \ac{msr} and \ac{cr} values for VDE and VRE, as detailed in Table~\ref{tab:combined-models}, corroborate the insights gained from our manual analysis in RQ2. These findings highlight the efficacy of our output control method in enhancing the likelihood of extracting potentially compilable source code from inference output, thereby boosting the \ac{cr}. In contrast, neglecting the output format and attempting direct compilation of inference outputs (VDE) achieves the lowest average \ac{cr} at just 7.95\%. This stark difference underscores the critical role of the output format and control when reporting and interpreting the performance of \acp{llm} for code translation. 

The three combinations (CRE, VDE, VRE) share similar patterns in terms of \ac{ca} with the other two metrics (\ac{msr}, \ac{cr}). The lowest \ac{ca} value is 4.92\% for VDE. The second highest \ac{ca} is for VRE, with 35.74\%. Even though the Vanilla Prompt is not designed to increase the chances of outputting a specific format, it shows higher \ac{cr} and \ac{ca} than VDE.

Additionally, it can be observed that Magicoder models and WizzardCoder 34B had a CR of 0\% in VDE, with no generation output that can be compiled. To further examine the possible causes of this phenomenon, we randomly sampled 350 generated outputs for each of WizzardCoder 34B, Magicoder-CL, and Magicoder-S-CL models. Our inspection confirms that all the 350 samples for each of the models contain \textit{Additional text} or an output format different than \textit{Direct Code}, which explains why none of the generated outputs are compilable.

Lastly, we can observe that the \ac{cr} of Mixtral 8x7B decreased from 53.51\% in VRE to 0\% on CRE, which is the opposite of what we would expect in CRE. Manual inspection confirms that this is because 100\% of the 350 generated out samples are in another format than the one expected by our regex, such as \textit{Re-Wraps Code} or others. This showcases the importance of carefully inspecting the output format of each model when comparing them in a benchmark.

\noindent \textbf{Finding 6: The consideration of output formats can significantly alter the outcomes when benchmarking various LLMs.} In the context of CRE, our analysis reveals a noteworthy trend: WizzardCoder Python 34B emerges as the top-performing model, boasting an average \ac{ca} of 43.85\% across all languages. This finding is particularly striking when juxtaposed with the performance of Magicoder-S-CL, a relatively smaller model with 7B parameters, which secures the second-highest \ac{ca} at 43.30\% across all models. This suggests that model size might not be the sole determinant of effectiveness in the code translation task, same as reported in prior work~\citep{wei_magicoder_2023}. Conversely, in the VDE scenario, where output control is ignored, a different pattern is observed. Here, both the CodeLlama Instruct 7B and 13B models demonstrate superior \ac{ca} compared to others. This variance in the measured performance underlines the significant impact that the output format and control can have when reporting the capabilities of different \acp{llm}. %

\begin{tcolorbox}[enhanced,width=\textwidth,size=fbox,drop shadow southwest,sharp corners]

\textit{RQ3 Summary:} 
The characteristics of the output format and its control significantly influence the \ac{ca} metric. The lowest average \ac{ca}, at 4.92\%, is achieved when the generated output from the \acp{llm} is compiled directly (VDE). The highest average \ac{ca} across all models, at 31.92\%, is obtained using the CRE approach. We conclude that careful inspection
and source code extraction based on the output format of the models is necessary to achieve comparable performance results.

\end{tcolorbox}

\section{RQ4: \rqfour} \label{sec:rq4}

In RQ3, we demonstrate how the output format and the inclusion of natural language can significantly affect the reported performance of \acp{llm} when using execution-based metrics, e.g., \ac{ca}. Text-oriented metrics, like BLEU and CodeBLEU for code translation, compare the generated source code against a reference commonly referred to as the ground truth or golden answer. Unlike \ac{ca}, which offers a binary evaluation (pass or fail for each example), text-oriented metrics provide a continuous measure, capturing the degree of similarity or preference for each generated translation output. Given the substantial differences between these two types of evaluation metrics, this RQ extends our analysis by examining the impact of directly evaluating the \ac{llm}'s inference output using these metrics, as opposed to extracting and isolating the source code to remove extraneous text, which we address through our proposed lightweight approach. We hypothesize that additional natural language alongside the source code would decrease the metrics' value, as they would be more different than the ground truth. For consistency in comparing execution-based metrics, as in RQ3, we concentrate on evaluating the outputs of open-source \acp{llm}.

\subsection{Approach}
\label{sec:approach}

To address RQ4, we use the extended dataset we derived from Pan et al.~\citeyear{pan2024lost}(as detailed in Section~\ref{sec:method:subsec:data}), along with the inference outputs from 11 open-source \acp{llm} generated in RQ3. Our objective is to investigate the impact of the output format on the CodeBLEU, BLEU and CrystalBLEU scores. For the outputs generated by each prompt in our study (Vanilla and Controlled Prompt), we evaluate if removing non-source-code tokens such as natural language text using our Regex (i.e., CRE and VRE setting) leads to an increase in the metrics in comparison to directly evaluating the inference output, which contains natural language (i.e., CDE and VDE setting).

The combinations of prompt and output processing methods considered in this RQ are similar to those in RQ3, with the addition of a new case, CDE:
\begin{enumerate}[leftmargin=*]
  \item \textbf{Vanilla + Direct Evaluation (VDE):} We utilize the Vanilla Prompt and directly evaluate the generated output. We do not perform any regex matching or source code extraction.
  \item \textbf{Vanilla + Regex (VRE):} We utilize the Vanilla Prompt and apply our regex to extract source code.
  \item \textbf{Controlled + Direct Evaluation (CDE):} We utilize the Controlled Prompt. We do not perform any regex matching or source code extraction.
  \item \textbf{Controlled + Regex (CRE):} We utilize the Controlled Prompt and apply our regex to extract source code.
\end{enumerate}

For this experiment, we exclude Mixtral, as we demonstrated in RQ3 that the regex is ineffective at extracting source code for this model. We include CDE, however, as it can be reliably evaluated using text-based metrics. In RQ3, we excluded CDE because its maximum possible Compilation Rate was limited to 6.6\% (i.e., 100\% - 93.40\% MSR), due to the controlled prompt format introducing triple backticks. Without regex-based code extraction, as is the case in CDE, these backticks would lead to compilation failures. When we compare the metrics (BLEU and CodeBLEU) computed on the extracted code (CRE, VRE) against the metrics calculated over the direct output from inference (CDE, VDE), we exclude points where the Regex does not match to ensure the same data points for both experiments (intersection set). For the differences observed in the metrics across the experiments, we report the significance with alpha 0.05 and Cliff's Delta effect size.

\subsection{Results} 

Table~\ref{tab:code-bleu-eval-vanilla} and Table~\ref{tab:code-bleu-eval-control} present the average CodeBLEU, BLEU and CrystalBLEU values for the evaluations conducted on ten open-source \acp{llm}, using the Vanilla Prompt and Controlled Prompt, respectively. Additionally, Table~\ref{tab:code_bleu_bleu_trends_cre_vre} shows the number of samples in the dataset where the metric increased after removing non-code tokens compared to evaluating the inference output. 

\begin{table}[]
\caption{Evaluation of CodeBLEU, BLEU, and CrystalBLEU Scores along with Cliff's Delta and Significance for Vanilla Prompt Experiments.}
\label{tab:code-bleu-eval-vanilla}
\resizebox{\textwidth}{!}{%
\begin{tabular}{lrrrrrrrrrrrr}
\hline
\multicolumn{1}{c|}{\multirow{2}{*}{Model}} & \multicolumn{4}{c|}{CodeBLEU} & \multicolumn{4}{c|}{BLEU} & \multicolumn{4}{c}{CrystalBLEU} \\ \cline{2-13} 
\multicolumn{1}{c|}{} & \multicolumn{1}{l|}{\textbf{VRE}} & \multicolumn{1}{l|}{\textbf{VDE}} & \multicolumn{1}{l|}{\textbf{Significance}} & \multicolumn{1}{l|}{\textbf{Cliff's Delta}} & \multicolumn{1}{l|}{\textbf{VRE}} & \multicolumn{1}{l|}{\textbf{VDE}} & \multicolumn{1}{l|}{\textbf{Significance}} & \multicolumn{1}{l|}{\textbf{Cliff's Delta}} & \multicolumn{1}{l|}{\textbf{VRE}} & \multicolumn{1}{l|}{\textbf{VDE}} & \multicolumn{1}{l|}{\textbf{Significance}} & \multicolumn{1}{l}{\textbf{Cliff's Delta}} \\ \hline
CodeLlama-13b-Instruct-hf & 0.26 & 0.12 & Yes & 0.52 & 0.11 & 0.10 & No & 0.01 & 0.06 & 0.06 & Yes & 0.04 \\
CodeLlama-34b-Instruct-hf & 0.29 & 0.28 & No & 0.02 & 0.09 & 0.06 & Yes & 0.14 & 0.05 & 0.03 & Yes & 0.08 \\
CodeLlama-7b-Instruct-hf & 0.29 & 0.26 & Yes & 0.10 & 0.11 & 0.08 & Yes & 0.16 & 0.06 & 0.04 & Yes & 0.05 \\
Magicoder-CL-7B & 0.31 & 0.27 & Yes & 0.14 & 0.11 & 0.10 & No & 0.02 & 0.05 & 0.05 & Yes & 0.01 \\
Magicoder-S-CL-7B & 0.29 & 0.27 & Yes & 0.08 & 0.10 & 0.06 & Yes & 0.18 & 0.05 & 0.02 & Yes & 0.08 \\
WizardCoder-1B-V1.0 & 0.29 & 0.29 & No & -0.03 & 0.10 & 0.07 & Yes & 0.15 & 0.04 & 0.03 & Yes & 0.05 \\
WizardCoder-3B-V1.0 & 0.30 & 0.31 & Yes & -0.06 & 0.10 & 0.05 & Yes & 0.28 & 0.05 & 0.02 & Yes & 0.10 \\
WizardCoder-Python-13B-V1.0 & 0.31 & 0.30 & Yes & -0.03 & 0.11 & 0.05 & Yes & 0.26 & 0.05 & 0.02 & Yes & 0.11 \\
WizardCoder-Python-34B-V1.0 & 0.32 & 0.30 & Yes & 0.06 & 0.12 & 0.07 & Yes & 0.20 & 0.06 & 0.03 & Yes & 0.08 \\
WizardCoder-Python-7B-V1.0 & 0.29 & 0.30 & Yes & -0.05 & 0.08 & 0.07 & Yes & 0.12 & 0.04 & 0.03 & Yes & 0.02 \\ \hline
\textbf{Average} & \textbf{0.30} & \textbf{0.27} & \textbf{---} & \textbf{0.08} & \textbf{0.10} & \textbf{0.07} & \textbf{---} & \textbf{0.15} & \textbf{0.05} & \textbf{0.03} & \textbf{---} & \textbf{0.06} \\ \hline
\end{tabular}%
}
\end{table}

\begin{table}[]
\caption{Evaluation of CodeBLEU, BLEU, and CrystalBLEU Scores along with Cliff's Delta and Significance for Controlled Prompt Experiments.}
\label{tab:code-bleu-eval-control}
\resizebox{\textwidth}{!}{%
\begin{tabular}{lrrrrrrrrrrrr}
\hline
\multicolumn{1}{c|}{\multirow{2}{*}{Model}} & \multicolumn{4}{c|}{CodeBLEU} & \multicolumn{4}{c|}{BLEU} & \multicolumn{4}{c}{CrystalBLEU} \\ \cline{2-13} 
\multicolumn{1}{c|}{} & \multicolumn{1}{c|}{\textbf{CRE}} & \multicolumn{1}{c|}{\textbf{CDE}} & \multicolumn{1}{c|}{\textbf{Significance}} & \multicolumn{1}{c|}{\textbf{Cliff's Delta}} & \multicolumn{1}{c|}{\textbf{CRE}} & \multicolumn{1}{c|}{\textbf{CDE}} & \multicolumn{1}{c|}{\textbf{Significance}} & \multicolumn{1}{c|}{\textbf{Cliff's Delta}} & \multicolumn{1}{c|}{\textbf{CRE}} & \multicolumn{1}{c|}{\textbf{CDE}} & \multicolumn{1}{c|}{\textbf{Significance}} & \multicolumn{1}{c|}{\textbf{Cliff's Delta}} \\ \hline
CodeLlama-13b-Instruct-hf & 0.30 & 0.30 & Yes & 0.52 & 0.11 & 0.11 & No & 0.01 & 0.06 & 0.06 & Yes & 0.01 \\
CodeLlama-34b-Instruct-hf & 0.29 & 0.30 & No & 0.02 & 0.11 & 0.04 & Yes & 0.14 & 0.06 & 0.02 & Yes & 0.12 \\
CodeLlama-7b-Instruct-hf & 0.30 & 0.30 & Yes & 0.10 & 0.11 & 0.11 & Yes & 0.16 & 0.06 & 0.06 & Yes & 0.02 \\
Magicoder-CL-7B & 0.30 & 0.30 & Yes & 0.14 & 0.10 & 0.10 & No & 0.02 & 0.05 & 0.05 & Yes & 0.00 \\
Magicoder-S-CL-7B & 0.30 & 0.29 & Yes & 0.08 & 0.10 & 0.10 & Yes & 0.18 & 0.05 & 0.04 & Yes & 0.01 \\
WizardCoder-1B-V1.0 & 0.29 & 0.29 & No & -0.03 & 0.10 & 0.06 & Yes & 0.15 & 0.05 & 0.02 & Yes & 0.08 \\
WizardCoder-3B-V1.0 & 0.30 & 0.30 & Yes & -0.06 & 0.11 & 0.06 & Yes & 0.28 & 0.06 & 0.03 & Yes & 0.09 \\
WizardCoder-Python-13B-V1.0 & 0.30 & 0.30 & Yes & -0.03 & 0.11 & 0.09 & Yes & 0.26 & 0.05 & 0.04 & Yes & 0.05 \\
WizardCoder-Python-34B-V1.0 & 0.31 & 0.31 & Yes & 0.06 & 0.12 & 0.07 & Yes & 0.20 & 0.06 & 0.03 & Yes & 0.09 \\
WizardCoder-Python-7B-V1.0 & 0.30 & 0.30 & Yes & -0.05 & 0.11 & 0.11 & Yes & 0.12 & 0.06 & 0.05 & Yes & 0.01 \\ \hline
\textbf{Average} & \textbf{0.30} & \textbf{0.30} & \textbf{---} & \textbf{0.08} & \textbf{0.11} & \textbf{0.08} & \textbf{---} & 0.15 & \textbf{0.06} & \textbf{0.04} & \textbf{---} & \textbf{0.05} \\ \hline
\end{tabular}%
}
\end{table}

\begin{table}[]
\caption{Percentage of data points that had an increase in the value for the respective metric compared to when the metric was calculated directly from the inference output.}
\label{tab:code_bleu_bleu_trends_cre_vre}
\resizebox{\textwidth}{!}{%
\begin{tabular}{lrrrrrr}
\textbf{Model} & \multicolumn{1}{c}{\textbf{Code BLEU (CRE)}} & \multicolumn{1}{c}{\textbf{BLEU (CRE)}} & \multicolumn{1}{c}{\textbf{CrystalBLEU (CRE)}} & \multicolumn{1}{c}{\textbf{Code BLEU (VRE)}} & \multicolumn{1}{c}{\textbf{BLEU (VRE)}} & \multicolumn{1}{c}{\textbf{CrystalBLEU (VRE)}} \\ \hline
CodeLlama-13b-Instruct-hf & 52.28\% & 59.08\% & 41.51\% & 80.77\% & 38.46\% & 57.69\% \\
CodeLlama-34b-Instruct-hf & 40.84\% & 80.01\% & 53.95\% & 52.48\% & 79.16\% & 55.78\% \\
CodeLlama-7b-Instruct-hf & 54.18\% & 64.09\% & 44.40\% & 61.0\% & 69.71\% & 44.40\% \\
Magicoder-CL-7B & 66.9\% & 62.37\% & 42.35\% & 75.39\% & 65.86\% & 43.27\% \\
Magicoder-S-CL-7B & 67.13\% & 62.41\% & 41.89\% & 54.32\% & 79.41\% & 54.51\% \\
WizardCoder-1B-V1.0 & 45.18\% & 75.2\% & 46.82\% & 55.84\% & 76.38\% & 46.75\% \\
WizardCoder-3B-V1.0 & 52.62\% & 75.68\% & 51.50\% & 43.09\% & 83.68\% & 54.38\% \\
WizardCoder-Python-13B-V1.0 & 53.3\% & 65.15\% & 46.73\% & 44.89\% & 81.14\% & 54.84\% \\
WizardCoder-Python-34B-V1.0 & 46.16\% & 76.26\% & 54.58\% & 54.75\% & 74.79\% & 52.59\% \\
WizardCoder-Python-7B-V1.0 & 65.31\% & 61.08\% & 41.89\% & 51.99\% & 72.26\% & 41.41\% \\ \hline
\textbf{Average Increase} & 54.39\% & 68.13\% & 46.56\% & 57.45\% & 72.08\% & 50.56\% \\ \hline
\end{tabular}%
}
\end{table}

\vspace{0.1cm}

\noindent \textbf{Finding 7: When non-source-code tokens are filtered out, the BLEU score increased on average 68.13\% and 72.08\% of the evaluated samples in CRE (using Controlled Prompt) and VRE (using Vanilla Prompt), respectively. In the case of CrystalBLEU, 46.56\% (CRE) and 50.56\% (VRE) of the samples increased their score. This difference is statistically significant for the majority of the models.} For example, as shown in Table~\ref{tab:code-bleu-eval-vanilla}, for the MagiCoder-S-CL-7B model, the average BLEU score with the Vanilla Prompt increased from 0.06 (VDE) to 0.10 (VRE), with a statistically significant difference and a small effect size (Cliff's Delta = 0.18). The degree of impact varies across models and prompts, as shown in Table~\ref{tab:code_bleu_bleu_trends_cre_vre}. For instance, the WizardCoder-3B-V1.0 model exhibited the largest improvement when the Vanilla Prompt was used. The BLEU score increased in 83.68\% of the cases after removing non-source-code tokens. When the Controlled Prompt was used, the CodeLlama-34b-Instruct-hf model exhibited an increase of BLEU in 80.01\% cases after removing non-source-code tokens, the highest among the ten models.

\vspace{0.1cm}
\noindent \textbf{Finding 8: The presence of non-source-code tokens complicates the interpretation of CodeBLEU. On average, filtering out these tokens increases CodeBLEU in 54.39\% of evaluated samples with Controlled Prompts (CRE) and 57.45\% with Vanilla Prompts (VRE). In the remaining cases, filtering either had no effect or decreased the CodeBLEU score.} Unlike BLEU, which is based on n-grams, CodeBLEU evaluates n-gram similarity and factors like syntax tree structure, data flow, and semantic match. The interplay between these components makes the metric more sensitive to variations in the structure, especially when additional text is included. For instance, we observe that DataFlow is the component with the highest decrease, 43.76\% in VRE and 31.24\% in CRE, after filtering out natural language, as seen in Fig.~\ref{fig:plot-control-variations} and Fig.~\ref{fig:plot-vanilla-variations}. This suggests that CodeBLEU's Data Flow component may be particularly affected by the inclusion of natural language, as the component decreased when removed, leading to lower scores.

\begin{figure}[!htb]
\centering
\includegraphics[width=0.87\textwidth]{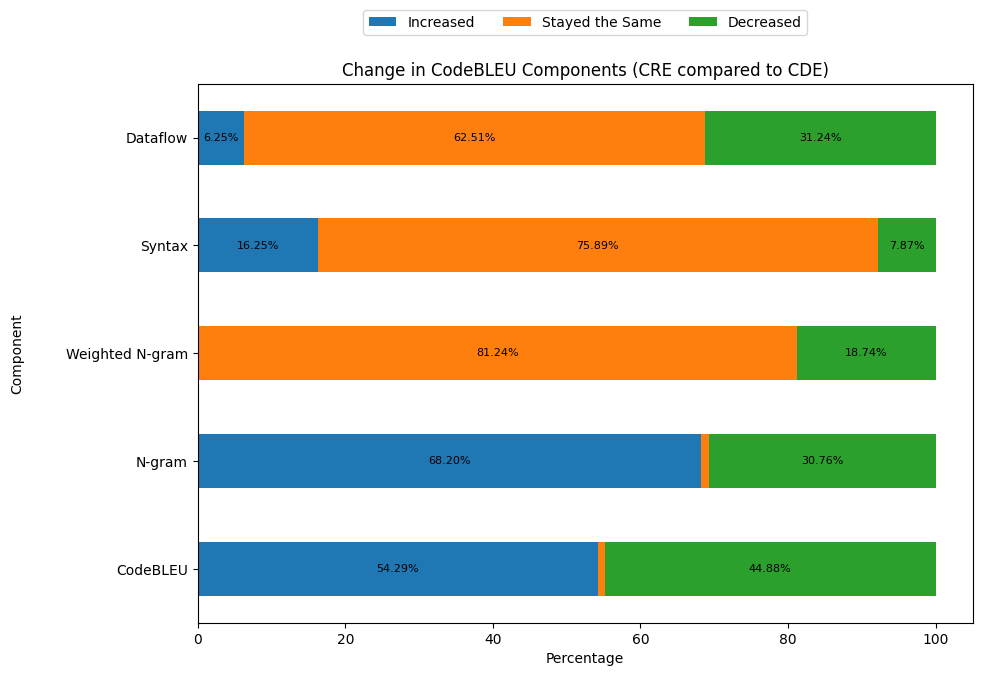}
\caption{Variations in the components of the CodeBLEU metric following the extraction of source code using regular expressions with the Control Prompt.} 
\label{fig:plot-control-variations}
\end{figure}

\begin{figure}[!htb]
\centering
\includegraphics[width=0.87\textwidth]{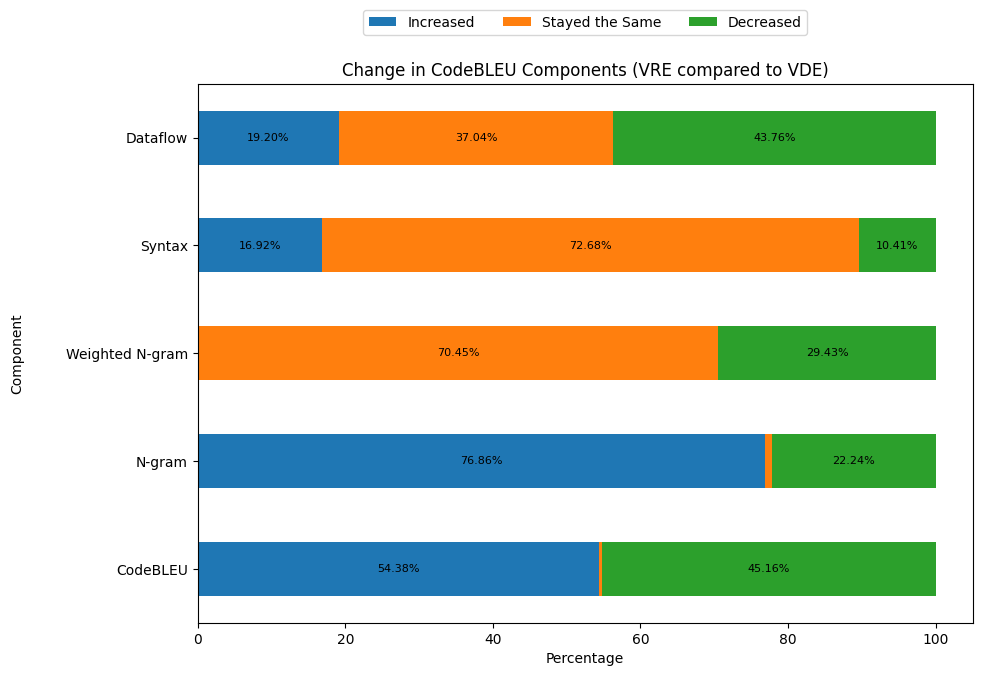}
\caption{Variations in the components of the CodeBLEU metric following the extraction of source code using regular expressions with the Vanilla Prompt.} 
\label{fig:plot-vanilla-variations}
\end{figure}

We manually reviewed 50 randomly sampled inference outputs from the Vanilla Prompt, where the Data Flow score decreased after extracting the source code using regular expressions. All of these outputs included natural language text, and in 46 out of 50 cases, the model added additional text describing the modifications it made during translation, with specific references to variables, functions, and other identifiers. This suggests that the presence of additional text significantly inflates the Data Flow component score. When this text is removed during extraction, the Data Flow score decreases, affecting the reliability of the CodeBLEU metric.

\begin{tcolorbox}[enhanced,width=\textwidth,size=fbox,drop shadow southwest,sharp corners]

\textit{RQ4 Summary:} BLEU and CodeBLEU metrics are influenced by the output format and non-source-code tokens in the inference output. When utilizing regular expressions to capture source code, the BLEU scores show a noticeable increase. Furthermore, Additional Text appears to inflate the Data Flow component of CodeBLEU artificially.
\end{tcolorbox}

\section{RQ5: \rqfive} \label{sec:rq5}

Open-source and closed-source \acp{llm} are widely used for coding tasks, including code translation. Closed \acp{llm} are often more effective at following instructions, which suggests they may be more likely to generate clean source code output when they understand this is the expected result for a code translation task (due to instruction fine-tuning and human alignment) or when explicitly directed to omit additional text. In this RQ, investigate whether closed models exhibit similar output format biases. Specifically, we aim to answer whether these models consistently produce source code in a format that is directly suitable for evaluation and if they present a distribution of output formats.

\subsection{Approach}

We selected five closed \acp{llm}: GPT-3.5, GPT-4, GPT-4o, GPT-4o mini, and Gemini 1.5 Flash (the rationale for choosing these models is discussed in Section~\ref{sec:method:subsec:data}). For this analysis, we reuse the subset created in RQ1 to analyze the output format for open models. Reusing this dataset allows us to maintain consistency in our evaluation, ensuring that the results can be directly compared with the open-source models assessed earlier.

We begin by using the Vanilla Prompt, which as shown in Fig. \ref{fig:prompts-used}, instructs the model to ``output code only'' when generating inference outputs. This allows us to observe how these closed models follow basic, straightforward instructions and assess the consistency of their output formats without additional control (e.g., the directives provided in the Controlled Prompt). We then manually examine the output formats generated by the four models to identify any inconsistencies.

In cases where multiple output formats are produced, we further investigate whether our controlled prompt can effectively standardize the outputs. This step helps us evaluate whether closed-source models have different output format distributions.

\subsection{Results} Fig.~\ref{fig:closed-models} presents the distribution of output formats observed in the inference outputs generated by the five selected closed \acp{llm}. Aligned with our findings on open-source \acp{llm}, we identified two distinct output formats: Wrapped Code and Direct Output. Additionally, we note that no Additional Text is output in any of the closed-source models, indicating they correctly followed the instructions in the prompt.

\begin{figure}[ht]
\centering
\includegraphics[width=\textwidth]{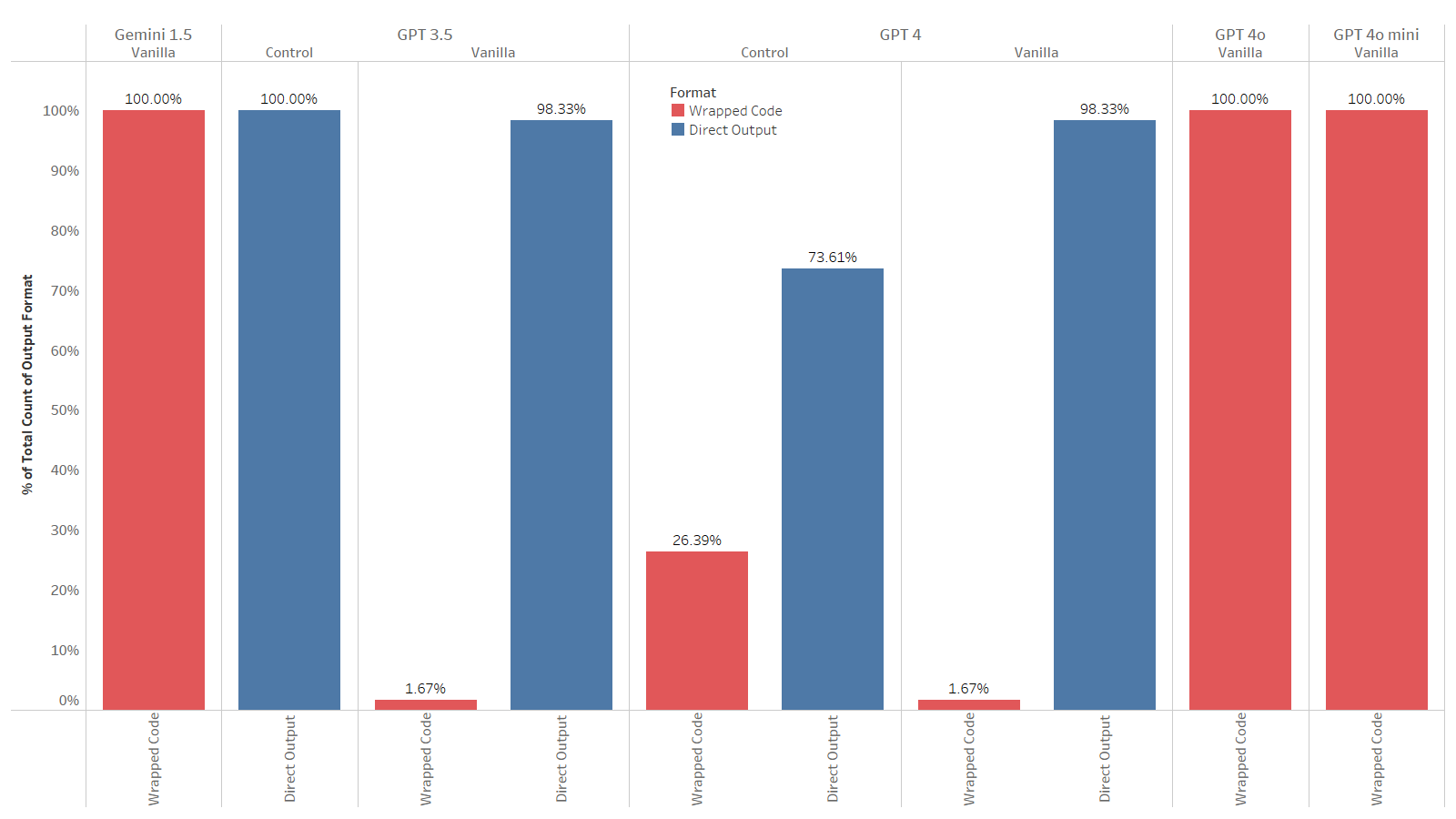}
\caption{Percentage of inference outputs for each model in their respective output format.} 
\label{fig:closed-models}
\end{figure}

\noindent \textbf{Finding 9: Closed \acp{llm} typically produce inference outputs with a consistent format within each model. However, the output format can vary significantly between models, even within the same model family.} As shown in Fig.~\ref{fig:closed-models}, when Vanilla Prompt is used, three out of the five analyzed models, Gemini 1.5 Flash, GPT-4o, and GPT-4o mini, consistently produce output exclusively in Wrapped-Code format, generating one coherent output format. In contrast, earlier iterations of GPT, i.e., specifically GPT-3.5 Turbo and GPT-4, predominantly favour Direct Output format using the Vanilla Prompt, with only a small fraction (1.67\%) of their outputs remaining in Wrapped-Code format.

Notably, attempts to influence the output format through prompt engineering yield inconsistent results. For instance, with GPT-3.5 using the Controlled Prompt leads to the Direct Output format increasing from 99.33\% to 100\%. However, in the case of GPT-4, this same approach leads to a decrease in Direct Output from 98.33\% to 73.61\% and an increase of Wrapped-Code format from 1.67\% to 26.39\%.

These findings indicate that understanding the output format distribution remains relevant in certain closed-source models, and a correct extraction method should be considered for fair comparisons across models. For instance, accounting for output format is crucial when benchmarking open-source models against closed-source ones, as the latter tend to exhibit more consistent output format characteristics, which could bias the results in their favor.

\begin{tcolorbox}[enhanced,width=\textwidth,size=fbox,drop shadow southwest,sharp corners]

\textit{RQ5 Summary:} 
Models such as GPT-4o, GPT-4o mini, and Gemini 1.5 Flash exhibit a consistent output format. This is very relevant given that they are often compared to open-source models that do not have a consistent output format, incrementing the chances of output format bias. For other models, such as GPT 3.5 Turbo and GPT-4, it is essential to develop effective strategies for managing output formats to minimize biases in evaluation. 
\end{tcolorbox}

\section{Discussion} \label{sec:disc}

In this section, we discuss the implications of our findings and offer recommendations for key stakeholders, namely, researchers benchmarking LLMs on coding tasks, particularly code translation, and developers who utilize LLMs in their day-to-day coding activities, including code translation.

\subsection{Causes of Output Format Bias}
\label{sec:causes-output-format}

Understanding the causes of output format bias is essential for mitigating its effects. However, fully addressing this question remains difficult due to the lack of transparency surrounding the training data of most large language models (LLMs). Although many open-source models provide access to their model weights, their training datasets are not publicly available, limiting our ability to identify contributing factors.

Magicoder is a notable exception, as it is the only model among those evaluated in our work with publicly available training data. Magicoder was instruct-tuned on the OSS-INSTRUCT dataset \citep{wei_magicoder_2023}, which consists of prompt-code pairs in the Wrapped Code format. This suggests that exposure to Wrapped Code formatted examples during training can influence a model’s tendency to produce consistently formatted outputs.

For closed-source models, such as those developed by OpenAI, the proprietary nature of their training pipelines prevents us from drawing conclusions. Nonetheless, we hypothesize that newer models like GPT-4o and GPT-4o-mini may been trained with output control in mind. As illustrated in Fig.~\ref{fig:closed-models}, earlier versions like GPT-3.5 and GPT-4 frequently defaulted to the Direct Output format and were less responsive to output control prompts. In contrast, GPT-4o and GPT-4o-mini natively produce code in the Wrapped Code format without requiring explicit prompting, which suggests a shift toward incorporating output control measures during the model construction.

\subsection{Mitigating Output Format Bias}

We propose a combination of prompt engineering and lightweight post-processing using regular expressions to control output formats. This approach has proven effective for open-source \acp{llm}, achieving an impressive MSR of 93.40\%, as shown in Table~\ref{tab:combined-models}. These techniques are not limited to benchmarking but can also be implemented as a post-processing step in practical code translation applications.

We offer tailored suggestions for closed \acp{llm} based on Finding 9. Specifically, for GPT-3.5, we recommend directly evaluating the inference outputs generated with the Control Prompt. In the case of GPT-4, regular expressions should be applied, and if no match is found, the inference output should be evaluated directly.

We also highlight the potential presence of internal mechanisms in certain models, such as Mixtral, that may affect the steerability of output formats. Researchers should be aware of such model-specific behaviors and adjust their code extraction methods accordingly, underscoring the need for customized approaches in the evaluation process.

\textbf{Overall, for both open-source and closed \acp{llm}, developing and employing effective strategies to control output formats is essential to minimize bias and ensure reliable evaluation of large language models in code translation.}

\subsection{Output Format Bias Mitigation in Practice}
\label{sec:mitigation-in-practice}

Since the publication of our earlier work \citep{macedo2024exploring}, subsequent research has incorporated our output control strategies, including recent efforts such as InterTrans \citep{macedo2024intertrans} and AlphaTrans \citep{ibrahimzada_alphatrans_2025}.

To better understand the practical impact of output format bias, we conduct an ablation study using AlphaTrans as a case study. AlphaTrans \citep{ibrahimzada_alphatrans_2025}, a repository-level framework for Java-to-Python code translation, adopts our recommendation of appending triple backticks to prompts in order to control the output format\footnote{\url{https://github.com/Intelligent-CAT-Lab/AlphaTrans/blob/593a428bb93031fe169f81109fca97b4cb93dedb/src/translation/prompt_generator.py\#L383}}. We replicate their experimental setup, using \textit{deepseek-coder-33b-instruct} \citep{deepseek2024deepseekcoder33b} with greedy decoding, and remove the triple-backtick suffix from the translation prompt templates. The results, shown in Table~\ref{tab:main_methods_parseable}, indicate that their approach suffers from a substantial decline in the percentage of syntactically valid Application Main Fragments (AMFs) after the removal of the output control. Further analysis on the translation logs confirms that this drop is primarily due to the model generating additional natural language, which hinders the reliable extraction of code from its outputs.

Importantly, all translated projects had some percentage of syntactically correct code without output format control, indicating that the absence of output format control does not universally result in failure. This observation aligns with our results and reinforces the importance of output format considerations for reliable evaluation. The AlphaTrans authors' proactive adoption of this technique underscores its practical relevance and importance.

\begin{table}[h!]
\centering
\caption{Percentage of Application Main Fragments (AMFs) with no syntax issues in AlphaTrans compared to AlphaTrans with no Output Control}
\label{tab:main_methods_parseable}
\begin{tabular}{lrrr}
\toprule
\textbf{Project} & \textbf{AlphaTrans} & \textbf{AlphaTrans No-Control} & \textbf{Diff} \\
\midrule
JavaFastPFOR & 95.32\% & 19.25\% & -76.07\% \\
commons-cli & 100.00\% & 18.68\% & -81.32\% \\
commons-codec & 98.53\% & 26.18\% & -72.35\% \\
commons-csv & 98.72\% & 12.77\% & -85.96\% \\
commons-exec & 100.00\% & 29.44\% & -70.56\% \\
commons-fileupload & 100.00\% & 30.73\% & -69.27\% \\
commons-graph & 99.63\% & 20.33\% & -79.30\% \\
commons-pool & 100.00\% & 29.47\% & -70.53\% \\
commons-validator & 99.23\% & 23.07\% & -76.16\% \\
jansi & 99.76\% & 28.61\% & -71.15\% \\
\bottomrule
\end{tabular}
\end{table}

\subsection{The Existence and Impact of Output Format Bias}
\label{label:sec:existence}

We analyzed the output formats of 11 open-source \acp{llm} and five closed \acp{llm}, identifying four distinct ways in which source code is presented in LLM inference outputs: Direct Output, Wrapped Code, Unbalanced Back-ticks, and Re-Wrapped Code. Additionally, in all open-source \acp{llm}, we found that source code was often mixed with additional text. This inconsistency in output formats can introduce bias when comparing models using execution-based or text-oriented evaluation metrics, which typically assume that the target is pure source code. Our first finding underscores this issue for open-source \acp{llm}, with post-processing required to extract the source code in up to 73.6\% of outputs. \textbf{\textit{Researchers and practitioners should not assume that LLMs will produce a single and consistent output format. Even with a fixed model, prompt, and dataset combination, there can be a variety of output formats.}}

Our analysis reveals several key issues that arise when output format bias is overlooked in the evaluation of \acp{llm} for code translation:
\begin{itemize}
    \item There is a significant difference in output format bias between closed and open-source \acp{llm}. While open-source \acp{llm} outputs often require substantial post-processing—even when the prompt explicitly instructs the model to generate only source code—most recent and widely used closed \acp{llm} tend to produce more consistent outputs, either within a single model or across multiple models. Consequently, the performance of open-source \acp{llm} may be underestimated if their outputs, though accurate, contain additional text or symbols that obscure the underlying source code.
    \item Output format bias can also be influenced by the design of the prompt, as highlighted in Finding 2. Different stakeholders may phrase their prompts in various ways, and this variation can introduce additional biases in model evaluation. If the output format is ignored and the direct output is used for evaluation instead of the extracted source code, it can lead to inaccurate assessments of model performance.
    \item Ignoring output format bias leads to lower reported performance or hard-to-interpret results. In our case studies (RQ3 and RQ4), we show how controlling the output format (i.e., mitigating output format bias) can significantly impact both execution-based and text-oriented metrics. Finding 5 highlights a notable improvement in CA, with the CRE combination yielding an average CA of 31.92\% across all models, compared to just 4.92\% when directly compiling the raw outputs from LLMs. Finding 8 reveals another challenge when output format bias is ignored in interpreting text-oriented metrics.
    \item Although closed \acp{llm} exhibit more consistent output formats compared to open-source models, their preferred output format can still vary between models, even within the same family (see Finding 9).
    \item While some recent closed-source models, such as GPT-4o, appear to avoid output bias issues, other recent models such as DeepSeek Coder can exhibit such biases, as demonstrated in Sec. \ref{sec:mitigation-in-practice}. This emphasizes that the findings of our study remain highly relevant regardless of specific model choices. Therefore, we urge researchers and practitioners not to assume that newer models are free from output format bias.
\end{itemize}
\textbf{Therefore, controlling the output format is crucial for the accurate evaluation of code translations. We emphasize the importance of considering both output format and extraction methods as key factors when assessing LLMs for code translation tasks. To ensure fair and reliable comparisons, we recommend that stakeholders report metrics such as MSR and CR, alongside their assumptions about the output format and their output format bias mitigation strategy, to ensure reliable and comparable evaluations of \acp{llm} for code translation.}

\section{Threats to Validity} \label{sec:threats}
\noindent\textbf{External Validity.}  We acknowledge several limitations that might impact the external validity of our findings. They are mainly introduced by the selection of the dataset, target LLMs, and considered prompts. Firstly, our research relies on a subset of the CodeNet benchmark. While this may limit the diversity of the programs (source code) and target PLs, CodeNet is a comprehensive and widely recognized benchmark in related work, and our selection of PLs is based on their popularity. Such a setting is also considered in a recent related work~\citep{pan2024lost}. This subset offers a diverse range of programming problems and solutions (we considered five popular programming languages and corresponding 20 PL translation cases), which we believe enhances the relevance and applicability of our findings. Regarding our model selection, we have considered 11 open-source LLMs and 5 closed LLMs from six families. Although this represents a constrained subset of available LLMs, these were carefully chosen for their popularity and recent advancements in the field. We choose instruct-tuned models rather than base models as instruct-tuned models are specifically optimized to follow instructions more effectively. This means they are more likely to produce outputs that are closely aligned with the given prompts.

Results in RQ3 show that our current specific prompt and regex may not apply to all models, which is particularly evident in the case of the Mixtral 8X7B model. While this highlights a limitation in the universality of the proposed specific output control methods across different models, our design method, i.e., how we analyze the output format and control it using a combination of prompt and regex, can be generalized and easily adapted to other models.

\vspace{0.1cm}
\noindent\textbf{Internal Validity.} The main threats to internal validity are introduced by the design of the prompt template and the labeling of output format types. The prompt templates used in our experiments were custom-designed, which may not represent the optimal choice for each model. However, we carefully crafted these prompt templates based on templates recommended on the models' official sites. Regarding the labeling of output formats, this task was undertaken by the first author of this paper. While this could introduce subjective bias or errors in labeling and taxonomy, we established clear, easily identifiable categories to minimize ambiguity and enhance consistency in our categorization process. %

\vspace{0.1cm}
\noindent\textbf{Construct Validity.} In this paper, as the CodeNet benchmark contains one test case per sample, there may be translations considered computationally accurate, but that do not contain identical functionality to the input program, i.e., false positives. This may influence the results reported in Table~\ref{tab:combined-models} on RQ3. However, our main goal is to point out the importance of considering output format control when utilizing LLMs for code translation, and such a conclusion can be supported by other metrics, such as extraction success rate and compilation rate, which are not sensitive to the quality of the test cases.

\section{Conclusion}\label{sec:conclusion} 
In this study, we conducted an empirical analysis of the output formats generated by 16 popular instruct-tuned large language models (LLMs), including both open-source and closed-source models, in code translation, using 3,820 translation pairs. Our findings reveal significant variability in the output formats produced by these LLMs, introducing a new type of bias, i.e., output format bias. Despite utilizing prompts that follow recommended strategies from official LLM documentation, we observed that the outputs frequently included additional text or presented code in quoted or partially quoted formats, negatively affecting the compilation success rate. To address this challenge, we proposed an output control approach that combines specific prompt designs with regular expressions to extract code cleanly from the outputs. Our case study, based on 3,820 translation pairs, demonstrates the effectiveness of this approach in extracting source code from LLMs' output and improving reported evaluation metrics.

Our research underscores the critical role of output format, prompt engineering, and code extraction methods in the evaluation of LLMs for code translation tasks. The variability in how different models generate output formats, along with their sensitivity to prompt design, can significantly impact key evaluation metrics—both execution-based and text-oriented. Ignoring these factors could lead to misinterpretations and unfair comparisons between LLMs. We offer recommendations for practitioners to accurately assess, control, and mitigate output format bias in LLM evaluations. Our insights contribute to improving benchmarking practices in code translation tasks and provide a framework that could extend to the evaluation of LLMs across other coding tasks, helping to better measure their potential and reliability.

\subsection{Funding} 
We express our gratitude to the Natural Sciences and Engineering Research Council of Canada (NSERC) for their support, with funding reference number RGPIN-2019-05071. Additionally, we extend our appreciation to the Vector Institute for its offering of the Vector Scholarship in Artificial Intelligence, which was awarded to the first author. The findings and opinions expressed in this paper are those of the authors and do not necessarily represent or reflect those of Huawei and/or its subsidiaries and affiliates.

\subsection{Data Availability Statements}
The results, source code, and data related to this study are available at \url{https://github.com/RISElabQueens/forge24-code-translation/tree/main}

\clearpage

\bibliographystyle{ACM-Reference-Format}

\bibliography{references}

\clearpage

\end{document}